%% file: scalar_new.tex
\documentclass[12pt]{article}
\makeatletter
\renewcommand\paragraph{\@startsection{paragraph}{4}{\z@}%
	{-2.5ex\@plus -1ex \@minus -.25ex}%
	{1.25ex \@plus .25ex}%
	{\normalfont\normalsize\bfseries}}
	\makeatother
\usepackage[symbol]{footmisc}
\renewcommand{\thefootnote}{\fnsymbol{footnote}}
\usepackage{amsmath}
\usepackage[fleqn]{mathtools}
\usepackage{slashed}
\usepackage{graphicx}
\usepackage{subfig}
\usepackage{hyperref}
\hypersetup{
	colorlinks=true,
    bookmarks=true,
    linkcolor=blue,
    filecolor=magenta,      
    urlcolor=blue,
    citecolor=blue,
}
\def\d{{\rm d}}

\def\gev2{\hbox{GeV}^2}
\def\<{\langle}
\def\>{\rangle}

 \usepackage[normalem]{ulem}
\usepackage{color}

\def\lapp{\mathrel{\rlap{\raise.5ex\hbox{$<$}}
                    {\lower.5ex\hbox{$\sim$}}}}
\def\gapp{\mathrel{\rlap{\raise.5ex\hbox{$>$}}
                    {\lower.5ex\hbox{$\sim$}}}}

\begin{document}
% \tableofcontents
\begin{flushright}
%November 1, 2017 \\
\today
\end{flushright}

\begin{center}
{\Large \bf
Infrared finiteness of a thermal theory of scalar electrodynamics to all
orders}\\ [0.2cm]
Pritam Sen\footnote[1]{pritamsen@imsc.res.in} and  D.
Indumathi\footnote[2]{indu@imsc.res.in} \\ [0.2cm]
The Institute of Mathematical Sciences, Chennai and
Homi Bhabha National Institute, Mumbai \\ [0.5cm]
Debajyoti Choudhury\footnote[3]{debajyoti.choudhury@gmail.com}\\[0.2cm]
Department of Physics and Astrophysics, University of Delhi,
Delhi 110 007, India \\ [1cm]
\end{center}

\renewcommand*{\thefootnote}{\arabic{footnote}}
\centerline{\underline{Abstract}}
\begin{quote}
  Models explaining dark matter typically include interactions with
  charged scalar and fermion fields. The Infra-Red (IR) finiteness of
  thermal field theories of charged fermions (fermionic QED) has been
  proven to all orders in perturbation theory. Here we reexamine the IR
  behaviour of charged scalar theories at finite temperature.
  Using the method of Grammer and Yennie, we identify and
  factorise the infra-red divergences to all orders in perturbation
  theory. The inclusion of IR finite pieces arising from the 4-point
  interaction terms of scalars with photon fields is key to the
  exponentiation. We use this in a companion paper to prove the IR
  finiteness of the corresponding thermal theory which is of relevance
  in dark matter calculations.

\end{quote}
% 95.35.+d % Dark matter

% Introduction
% \label{sec:intro}
\input{Sections/sec_intro.tex}

% Real-time formulation of thermal field theory
\input{Sections/sec_thermal.tex}

% \input{Sections/sec_thermal_debu.tex}

% The IR behaviour of thermal scalar QED
\input{Sections/sec_scalar_ir.tex}

% \input{Sections/sec_scalar_ir_debu.tex}

% Summary and Discussion
\input{Sections/sec_concl.tex}

\appendix
\renewcommand{\theequation}{\thesection.\arabic{equation}}
\input{Sections/appa.tex} % Feynman rules
\input{Sections/appb.tex} % Feynman identities
\input{Sections/appc.tex} % virtual K photon insertions
\input{Sections/appd.tex} % virtual G photon insertions

\end{document}

%% file: Sections/sec_intro.tex
\section{Introduction}
\label{sec:intro}

At zero temperature, Bloch and Nordsieck \cite{BN} were among the first
to study the infra-red (IR) behaviour of fermionic QED. Later, it was
shown \cite{Low} that the cross section for the bremsstrahlung of very
low energy quanta in elementary particle collisions has an IR divergence:
\begin{equation}
\sigma_{\rm brems} = \frac{\sigma_0}{k} + \sigma_1 + k \sigma_2 +
\ldots~,
\end{equation}
where $k$ is the energy of the photon and $\sigma_j$ have appropriate
dimensions. It was further shown that $\sigma_0$ and $\sigma_1$ can be
calculated from the corresponding elastic amplitude for both scalar
and spinor cases at the leading order in perturbation theory,
calculated up to ${\mathcal O}(k)$.  This was later extended \cite{YFS}
for pure fermionic QED where it was shown that the (logarithmic) IR
divergences cancel to all orders (rendering the total cross section IR
finite) when both virtual and real photon emission corrections are
included. Such soft real emissions need to be included due to finite
detector resolution since they cannot be distinguished from the
virtual lower order process. Some of the technical shortcomings of
Ref.~\cite{YFS} such as translational and gauge-invariance were
addressed in a subsequent paper by Grammer and Yennie \cite{GY}.

Many clarifications and simplifications occurred over the next
decades, including \cite{Faddeev} the question about whether a charge
particle exists relativistically due to the IR structure of gauge
theories where the Green functions for charged matter have no poles but
a branch cut. This implies a soft cloud always surrounds each physical
charge. This question was addressed (positively) in
Ref.~\cite{Bagan,Horan} where they used velocity-superselection rules
inspired by heavy quark effective theory for abelian theories to
obtain on-shell Green's functions that are IR finite to all orders in
perturbation theory. Specifically, they used scalar QED for
simplicity, since Low~\cite{Low} had shown that the electron spin
structure does not affect the IR divergence as long as the matter
fields are massive. (The spin structure of massless QED makes its
asymptotic dynamics richer; for instance, collinear divergences turn
on.)  Scalar QED has also been studied recently \cite{Laddha} in the
context of its asymptotic symmetries and relation to Weinberg's soft
photon theorem.

Many papers have also addressed the IR finite remainder in such scalar
theories. For instance, in Refs.~\cite{Matsson1,Matsson2,Matsson3},
the factorisation and exponentiation of IR divergences is shown in a
translation and gauge-invariant way, using order-by-order agreement
with Operator Product Expansion (OPE) {\em before} summation and by
requiring that the exponentiation of all factorisable parts is done
{\em before} the integrations are carried out. Then the IR finite
remainder is defined in terms of correlations with respect to the
photon momenta in the integrands. This involves an all order
generalisation of Low's theorem and also includes a calculation of
both soft and hard photon contributions.

In the case of thermal field theory, there are additional {\em linear}
divergences owing to the nature of the thermal photon propagator. The
infra-red finiteness of such thermal QED with purely charged fermions has
been shown \cite{Indu,Weldon} to all orders in the theory. In particular,
both absorption and emission of photons with respect to the heat bath are
required \cite{Indu,Sourendu} in order to cancel the linear divergences
as well as the logarithmic subdivergences.

In the first of this set of two papers we address the proof of the
infrared finiteness, to all orders, of a thermal field theory of
pure charged scalars, referred to as scalar QED. In the second paper,
we apply these results, and the earlier results on the IR finiteness
of thermal fermionic QED, to address the issue of IR finiteness of of
thermal models of dark matter, thereby extending the results obtained in
Ref.~\cite{Beneke} at NLO to all orders.  The analysis is an extension
of that presented in Ref.~\cite{Indu} which was based on the approach
developed by Grammer and Yennie (GY) \cite{GY} and is motivated
by the results of Ref.~\cite{Beneke}. The crux of this paper is the
identification of the correct set of terms that allows the factorisation
and exponentiation of the IR divergent terms to all orders for thermal
scalar fields.

In contrast to fermionic QED, we now have not only the 3-point
scalar-photon-scalar vertex, but also 4-point (2-scalar-2-photon)
ones. These contribute through both {\em seagull} and {\em tadpole}
diagrams; see vertex diagrams in Appendix~\ref{app:Ascalar}. While the
result we obtain is similar to that obtained in the usual fermionic QED,
the inclusion of the seagull and tadpole diagrams give rise to additional
terms that are essential in order to achieve the exponentiation and
cancellation of IR divergent terms between real and virtual contributions.

In the next paper, Paper II, we apply our results to show the IR
finiteness of the corresponding thermal field theory of dark matter to
all orders.  This result is, thus, a generalisation of Ref.~\cite{Indu}
to include both charged fermions and scalars. Again, the key fact used
in the proof is that both photon absorption and emission diagrams are
required to cancel the linear sub-divergences. As mentioned earlier,
this was also noticed in the NLO calculation in Ref.~\cite{Beneke},
where the finite term has also been calculated to NLO.

In Section~\ref{sec:sQED}, we briefly review the propagator and vertex
structure of the relevant thermal field theory; details are given in
Appendix \ref{app:Ascalar}. We also review the approach of Grammer
and Yennie (GY) to address the IR behaviour of such field theories. In
Section~\ref{sec:scalar_ir}, we analyse the photon--scalar interactions
using the approach motivated by GY: by rearranging the polarisation
sums of the inserted virtual photons into so-called $K$-photon and
$G$-photon parts (see Eq.~\ref{eq:gammaprop}). This was used by GY
to establish the IR finiteness of fermionic QED to all orders. As in
the case of fermionic QED, the $K$-photon contributions are divergent;
however, in scalar QED, they can be factorised and exponentiated only
on inclusion of the additional vertices. In particular, the ${\mathcal
O}(k^2)$ IR finite contribution from the tadpole diagrams cancels a
similar contribution from the 3-particle interaction terms and enables
the factorisation. As far as we are aware, this observation of the
need for inclusion of the IR finite tadpole contributions in order
to achieve the factorisation and subsequent exponentiation of the IR
divergent parts, has not been pointed out in the literature before. The
$G$-photon contributions are finite, again, as was shown to be the case
for fermionic QED. Proof of IR finiteness of the $G$-photon insertions
is non-trivial due to the presence of both 4-point vertices as well as
thermal indices and is the second main contribution of this work. In
Section \ref{sec:scalar_ir}, the corresponding analysis for insertion
of real photons is also considered. A similar rearrangement of the
polarisation sums of real photons into $\widetilde{K}$ and $\widetilde{G}$
enables the IR divergent parts to be collected into the $\widetilde{K}$
contribution. We show that the IR divergent parts will cancel between the
virtual and real diagrams, that is, between the $K$ and $\widetilde{K}$
contributions, when they are added, order by order, in the theory. This
is achieved only when both real photon emission into, and absorption
from, the heat bath is taken into account. This establishes the infrared
finiteness to all orders for scalar thermal QED.  We end with some remarks
and discussion in Section~\ref{sec:concl}. Many technical details are
relegated to the appendices.

Appendix~\ref{app:Ascalar} lists the relevant Feynman rules while
Appendix~\ref{app:fidentities} lists some useful generalised Feynman
identities of use in the thermal field theoretic analysis.
The details of the calculation for the insertion of a virtual $K$ photon
into a lower order graph is found in Appendix~\ref{app:vkgamma} where it
is shown that the total contribution from all possible virtual $K$
photon insertions into an $n^{\rm th}$ order graph is a single term
proportional to the lower order matrix element itself (and is also IR
divergent). Details of the result for the insertion of a virtual $G$
photon in all possible ways into an $n^{\rm th}$ order graph is found
in Appendix~\ref{app:vggamma}; it is shown that all such virtual $G$
photon contributions are IR finite.

%% file: Sections/sec_thermal.tex
\section{Real-time formulation of thermal field theory}
\label{sec:sQED}

We review briefly the real-time formulation of thermal (scalar and photon)
fields in equilibrium with a heat bath at temperature $T$.  In the case
of such a thermal field theory, there is an additional complication
which can be understood in a real-time formulation \cite{realtime}
where the integration in the complex time plane is over a contour that
includes the temperature, chosen so that correct thermal averages of
the $S$-matrix elements \cite{Rivers} are obtained. The fields satisfy
periodic boundary conditions,
\begin{equation}
\varphi(t_0) = \varphi(t_0 - i \beta)~,
\end{equation}
where $\beta = 1/T$, with $T$ being the temperature of the heat bath. This
results in the well-known field-doubling, so that fields are of type-1
(physical) or type-2 (ghosts), with propagators acquiring $2 \times 2$
matrix forms. Only type-1 fields can occur on external legs (as mandated
by unitarity) while fields of both types can occur on internal legs,
with the off-diagonal elements of the propagator allowing for conversion
of one type into another.

Both scalar and photon field propagators assume matrix forms (see
Appendix~\ref{app:Ascalar} for details) with the (11) and (22) terms
having both $T=0$ and finite temperature contributions. In particular,
the photon propagator corresponding to a momentum $k$ can be expressed
(in the Feynman gauge) as,
\begin{align}
i {\cal D}^{ab}_{\mu\nu} (k) & = - i g_{\mu\nu} \, D^{ab} (k)~, 
\label{eq:thermalD}
\end{align}
where the information on the field type is contained in $D^{ab} (k)$;
see Appendix~\ref{app:Ascalar} for its definition.

Finally, the vertices, both 3-point and 4-point ones, are modified in
the thermal theory. Details are again in Appendix~\ref{app:Ascalar};
we only note here that all the fields at a given vertex {\em must be}
of the same type.

\subsection{The GY approach to study the IR behaviour}

Several methods can be adopted to prove all-order finiteness. For example,
one may consider propagators dressed with arbitrary coherent states. We
shall, instead, adopt a simpler method that lends itself more readily to
an understanding of the issues involved. The approach of GY, which we use
here, addressed the IR finiteness of fermionic QED at zero temperature,
and we extend this to a theory of charged scalars in contact with a heat
bath. GY started with an $n^{\rm th}$-order graph with $n$ photon-fermion
vertices and considered the effect of adding an additional real or virtual
photon to it. Since the photon is a boson, all symmetric permutations,
{\em i.e.}, all possible insertions, must be considered. In particular,
for the virtual photon insertion, they found it useful to express the
photon propagator as,
\begin{align} \nonumber
-i \frac{g_{\mu{\nu}}}{k^2+i\epsilon} & = \frac{-i}{k^2+i\epsilon} \,
		\left[\left(g_{\mu\nu} - b_k (p_f, p_i)k_\mu k_\nu
		\right) + \left(b_k (p_f, p_i) k_\mu k_\nu \right)
		\right]~, \nonumber \\
 & \equiv \frac{-i}{k^2+i\epsilon} \, \left[ G_{\mu\nu} +
 	K_{\mu\nu} \right]~.
\label{eq:gammaprop}
\end{align}
Here, $b_k$ depends on the momenta $p_f$, $p_i$, where the final and
initial vertices are inserted (and also implicitly on the momentum $k$
of the inserted $(n+1)^{\rm th}$ photon), and is defined such that the
so-called $G$-photon terms in the matrix element with $(n+1)$ photons are
IR finite (in both the $T=0$ and $T\ne 0$ cases for fermionic QED)
and the $K$-photon terms contain all the IR divergent terms:
\begin{align}
b_k(p_f,p_i) = \frac{1}{2} \left[ \frac{(2p_f-k) \cdot
(2p_i-k)}{((p_f-k)^2-m^2)((p_i-k)^2-m^2)} + (k \leftrightarrow -k)
\right]~.
\label{eq:bk}
\end{align}
Note that, on account of its $k$ dependence, $b_k$ does not represent
a gauge transformation.

On expressing the $(n+1)^{\rm th}$ virtual photon contribution in this
way, the $K$ photon contribution turns out to be proportional to the
matrix element of the underlying graph with $n$-photon vertices and has
a simple structure. The object of this paper is to obtain an analogous
result for a theory of thermal charged scalars.

Note that the factor $g_{\mu\nu}$ occurs in all components of the thermal
photon propagator, enabling a separation into $K$- and $G$-type photons,
just as before, with the same\footnote{Slightly different from that
used by GY, this definition is more suitable for thermal field theory
\cite{Indu}.} definition for $b_k$ as in Eq.~\ref{eq:bk}. We can therefore
apply the technique of GY to the case of thermal fields in equilibrium
with a heat bath at temperature $T$. There are two major differences
in this case, firstly, that the relevant part of the thermal photon
propagator is proportional to,
\begin{align}
i {\cal{D}}^{ab} (k) \sim &
	\left[ \frac{i}{k^2+i\epsilon} \delta^{ab} \pm  2 \pi
	\delta(k^2) N(\vert k^0 \vert)
	D^{ab}_{T} \right]~,
\label{eq:thermalprop}
\end{align}
where the first term corresponds to the $T=0$ contribution and the second
to the finite temperature part. The bosonic number operator in the second
term contributes an additional power of $k$ in the denominator in the
soft limit, since
\begin{equation} 
N(\vert k^0 \vert)  \equiv 
\frac{1}{\exp^{\vert k^0 \vert/T}  - 1} 
\; \stackrel{k \to 0}{\longrightarrow} \; \frac{T}{\vert k^0 \vert}~.
\end{equation}
Hence, it can be seen that the leading IR divergence in the finite
temperature part is linear rather than logarithmic as was the case at zero
temperature. Consequently, there is a residual logarithmic subdivergence
that must also be shown to cancel at finite temperatures, thus making
the generalisation to the thermal case non-trivial.

Secondly, it turns out that the inclusion of thermal matter fields
adds another layer of complexity to the analysis, since not only is
the propagator structure now different from the zero
temperature case, but, in contrast to the case of fermions, the number
operator corresponding to charged scalars is bosonic and hence can
potentially give rise to divergences as well.

In summary, the major differences between this and the earlier works
are as follows.
\begin{enumerate}
\item The scalar theory has additional vertices, including
the 4-point seagull vertices; see Fig.~\ref{fig:feynman} in
Appendix~\ref{app:Ascalar}. This contributes additional terms to both
the $K$ and $G$ photon insertions compared to the thermal theory with
fermions only.

\item The thermal theory has additional field types; in particular,
the thermal charged scalar legs add more complications compared to the
results with thermal fermions.
\end{enumerate}

We consider both modifications when analysing the IR behaviour of
thermal scalar QED in the next section.

%% file: Sections/sec_scalar_ir.tex
\section{The IR behaviour of thermal scalar QED}
\label{sec:scalar_ir}

In view of the discussion in the preceding section, we begin by
considering pure scalar QED, discounting quartic scalar
self-couplings\footnote{While such self-couplings do indeed exist in
the generic case (and definitely so for the squarks and sleptons,
entities that we shall be interested in, in the companion paper),
given the rather large masses of such scalars, these couplings would
play virtually no role in the processes of interest.}.  Thus, it
behoves us to start with the fundamental hard scattering process here,
{\em viz.}  $\gamma^{(*)} + \phi^{(*)} \to \phi^{(*)}$ where any of
the three lines could represent either an on-shell or an off-shell
particle. Higher order contributions would arise from the inclusion
of both virtual as well as soft real photons.

We begin by considering insertions on an $n$-photon graph with trilinear
(scalar-scalar-photon) vertices alone so that the $n$ vertices imply $n$
scalar--photon interactions (with the understanding that both vertices
of an internal line are counted). We will, subsequently, extend the
analysis to graphs with an arbitrary admixture of 3-point and 4-point
vertices.

\begin{figure}[htp]
\begin{center}
\includegraphics[width=0.7\textwidth]{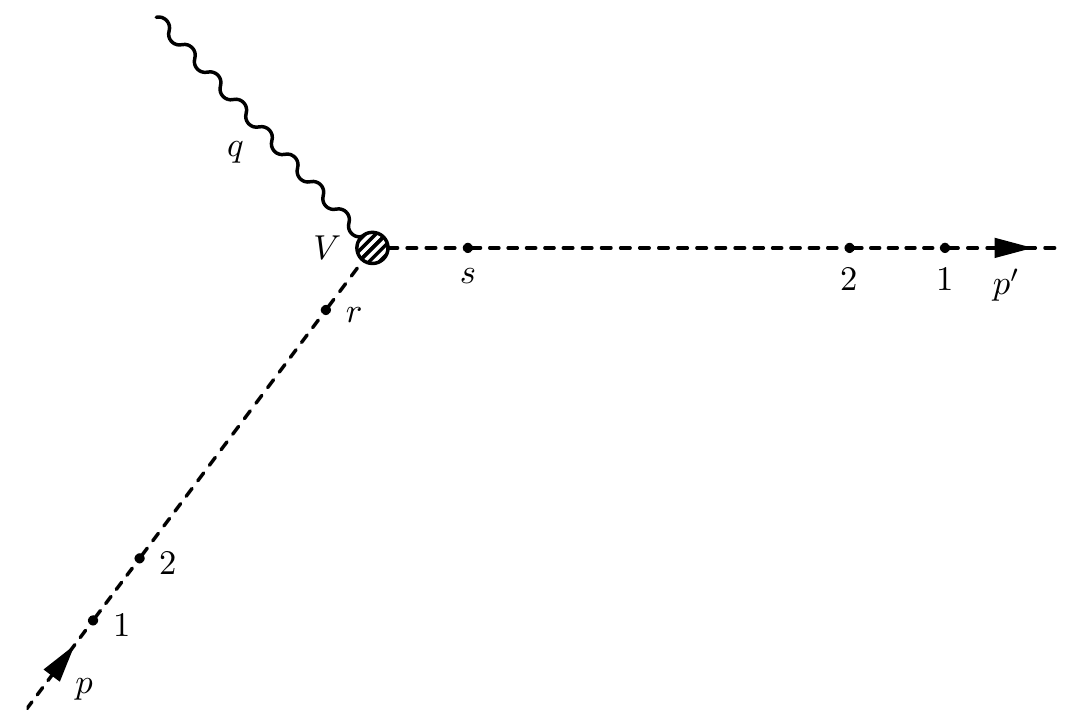}
\caption{\em Schematic of an $n^{\rm th}$ order graph of $\gamma^* \phi \to
\phi$, with $s$ vertices on the $p'$ leg and $r$ on the $p$ leg,
$r+s=n$. $V$ labels the special but arbitrary hard photon--scalar vertex.}
\label{fig:B1}
\end{center}
\end{figure}

The $(n)^{\rm th}$ order graph has $n$ trilinear vertices with $s$
vertices on the final scalar leg with 4-momentum $p'$ and $r \, (=
n - s)$ vertices on the initial scalar leg with 4-momentum $p$ (see
Fig.~\ref{fig:B1}). For reasons that will become clear later, these
vertices are already symmetrised. The photons carry away momentum $l_q,
\, q = 1, \cdots, s$, from the vertex $q$ on the $p'$-leg and momentum
$-t_q, \, q = 1, \cdots, r$, from the vertex $q$ on the $p$ leg. The
notation is arbitrary since the momenta may be entering or leaving the
vertex and the corresponding photon may be a real or virtual one.

Hence the momentum of the particle to the {\em right} of the $q^{\rm
th}$ vertex on the $p$ leg is $(p + \textstyle\sum_{i=1}^q t_i)$ while the
momentum corresponding to the particle line to the {\em left}
of the $q^{\rm th}$ vertex on the $p'$ leg is $(p' + \textstyle\sum_{i=1}^q
l_i)$.

In contrast to the fermionic case, which has only three-point vertices,
scalar QED admits of 4-point vertices as well, so that an additional
photon can be inserted at a new vertex (giving rise to a new 3-point
vertex) or at an already existing 3-point vertex, thus converting it to
a 4-point vertex. Thus the consideration of charged scalars requires
consideration of both types of insertions. This is true for both real
and virtual photon insertions.

We begin by considering insertion of an additional virtual photon.
Adopting the expression of $g_{\mu\nu}$ in the photon propagator in
terms of $K_{\mu\nu}$ and $G_{\mu\nu}$ (as in Eq.~\ref{eq:gammaprop}),
we start with the insertion of virtual $K$-photons (which are expected
to contain the IR divergent contributions) leaving the inclusion of the
$G$-photons (expected to give IR finite contributions) until later.

\subsection{Insertion of virtual $K$ photons}

\begin{figure}[h]
\includegraphics[width=\textwidth]{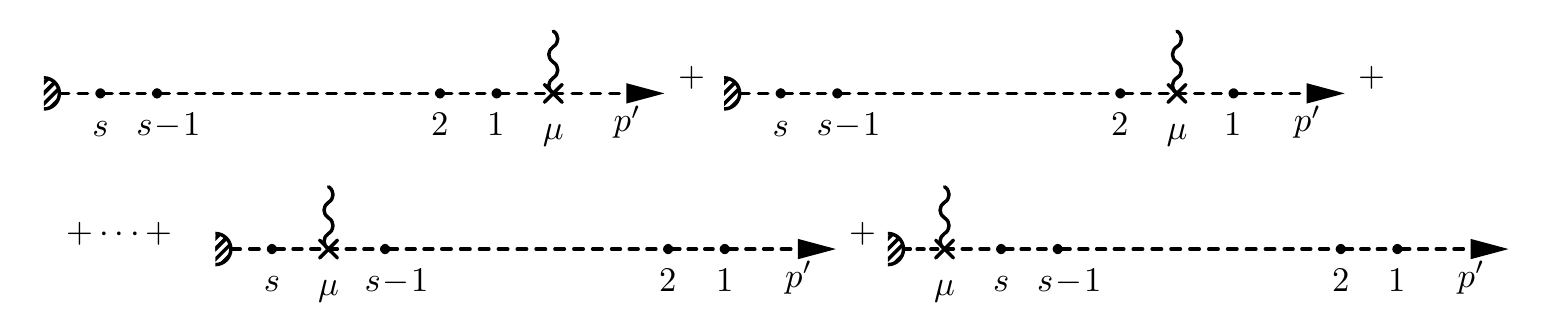}
% \caption*{Figs.~B.2}
\caption{\em Set of $(s+1)$ diagrams showing all possible trilinear
insertions of a virtual photon at vertex $\mu$ on the $p'$ leg of
a scalar/fermion.}
\label{fig:B2}
\end{figure}

\begin{figure}[h]
\includegraphics[width=\textwidth]{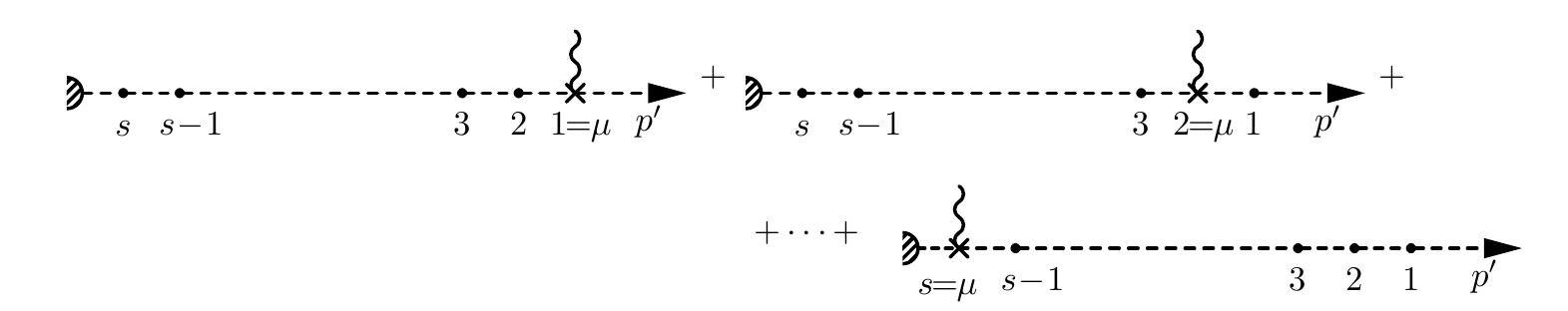}
% \caption*{Figs.~B.3}
\caption{\em Set of $s$ diagrams showing all possible insertions of a
virtual photon at vertex $\mu$ which is one of the already existing $s$
vertices on the $p'$ leg of a scalar particle, thus giving rise to
a 4-point vertex. Analogous diagrams for fermions do not exist.}
\label{fig:B3}
\end{figure}

Consider the insertion of one of the virtual $K$ photon vertices,
say $\mu$, on an external line. As per the Feynman rules listed in
Appendix~\ref{app:Ascalar}, there can be two types of vertices, with
one or two photon lines at each vertex, corresponding to 3-point or
4-point vertices respectively. (In addition, these fields carry a thermal
index, $t_a (=1,2)$, depending on the field type at the $a^{\rm th}$
vertex). Hence, there are two types of $K$ photon insertions possible;
one where the insertion is at a new vertex, forming a new 3-point vertex,
or one where the $(n+1)^{\rm th}$ $K$ photon is inserted on an already
existing vertex, thus forming a 4-point vertex. The total set of all
possible insertions of the $(n+1)^{\rm th}$ $K$ photon on the $p'$
line can be grouped into sets having the new $\mu$ vertex as a 3-point
or 4-point vertex, as shown in Figs.~\ref{fig:B2} and \ref{fig:B3}
respectively. In contrast, note that only the set of graphs shown in
Fig.~\ref{fig:B2} contributes if the $p'$-leg is a fermion line.

% \input Sections/fig_b2.tex
% \begin{figure}[htp]

% \input Sections/fig_b3.tex
% \begin{figure}[htp]

\begin{figure}[h]
\includegraphics[width=\textwidth]{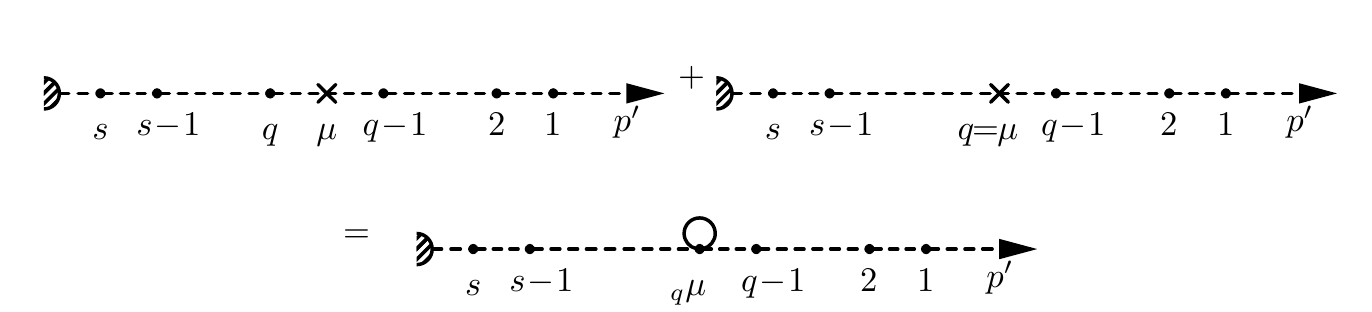}
% \caption*{Fig.~B.4}
\caption{\em Combining the two possible sets of insertions (as in
Figs.~\ref{fig:B2} and \ref{fig:B3}) of the $(n+1)^{\rm
th}$ virtual $K$ photon at the vertex $\mu$ on the $p'$ leg to give a single
{\em circled vertex}, ${}_q\mu$; see text for details. The photon lines
have been suppressed for clarity.}
\label{fig:B4}
\end{figure}

It is convenient to group 3- and 4-point vertices to obtain ``circled
vertices": for instance, consider the insertion of the $\mu$ vertex
to the {\em right} of a generic vertex $q$ or {\em at} the vertex
$q$. The corresponding two diagrams are shown in Fig.~\ref{fig:B4}
and the contribution from the sum of these is shown in the figure as a
circled vertex and denoted by ${}_q\mu$.

% \input Sections/fig_b4.tex
% \begin{figure}[hbt]

In the thermal case, the propagators contain more than just the
$1/(P^2-m^2)$ part and appear more complex. However, they satisfy
generalised identities, analogous to the zero temperature case, as shown
in Appendix~\ref{app:fidentities}, which can be used to simplify and
factor these contributions to obtain a similar result. Retaining only
the $k_\mu$ factor in the $(b_k k_\mu k_\nu)$ part of the $K$ photon
propagator (the $k_\nu$ factor will be similarly included when the other
vertex $\nu$ is inserted on the $p$ leg, and $b_k$ is an overall factor),
and omitting the other terms in the photon propagator for clarity, we
have (denoting a scalar propagator from the vertex $\mu_b$ of thermal
type $t_b$ to the vertex $\mu_a$ having fields of thermal type $t_a$
as $i{\cal S}^{t_at_b}(p'+\sum_{i=1}^q l_i,m) \equiv i
S^{ab}_{p'+\sum_q}\hbox{)}$,
\begin{align}
% \begin{split}
{\cal M}_{n+1}^{q~{\rm to~left~of~}\mu} & = \begin{multlined}[t]
 e^{s+1}(-1)^{\hbox{(}\sum_{i=1}^s t_i\hbox{)}+s} \cdots
       (-1)^{t_\mu+1} \left[ S\strut^{q-1,\mu}_{p'+\sum_{q-1}}
       \times ((2p'+2\Sigma_{q-1}+k)\cdot k) \times \right. \\
  \qquad \qquad \shoveright{\left.
  S\strut^{\mu,q}_{p'+\sum_{q-1}+k} \right]
  \left(2p'+2\Sigma_{q-1}+2k+l_q\right)_{\mu_q}
 S\strut^{q,q+1}_{p'+\sum_q+k}\cdots ~,} \nonumber
  \end{multlined} \\
% \end{split} \\
% \begin{split}
 & = e^{s+1}(-1)^{\hbox{(}\sum_{i=1}^s t_i\hbox{)}+s} \cdots 
\left[ S\strut^{q-1,q}_{p'+\sum_{q-1}} \delta_{t_\mu,t_q} -
S\strut^{q-1,q}_{p'+\sum_{q-1}+k} \delta_{t_\mu,t_{q-1}} 
\right] \cdots ~,
% \end{split} \\
\label{eq:qleftmu}
\end{align}
\begin{align}
{\cal M}_{n+1}^{q=\mu} & = e^{s+1}
	(-1)^{\hbox{(}\sum_{i\ne q}^s
	t_i\hbox{)}+s-1} \cdots (-1)^{t_\mu+1} \left[
	S\strut^{q-1,\mu}_{p'+\sum_{q-1}}
	(-2k_{\mu_q}) \delta_{t_\mu,t_q} \times
	S\strut^{q,q+1}_{p'+\sum_q+k} \right] \cdots ~; \nonumber \\ 
 & = e^{s+1}(-1)^{\hbox{(}\sum_{i=1}^s t_i\hbox{)}+s} \cdots 
\left[ S\strut^{q-1,q}_{p'+\sum_{q-1}} \delta_{t_\mu,t_q} 
	(-2k_{\mu_q}) 
	S\strut^{q,q+1}_{p'+\sum_q+k} \right] \cdots ~.
\label{eq:q=mu}
\end{align}
Here, $t_i (=1,2)$ denote the thermal indices of the inserted photons
and $t_\mu$ is the thermal index\footnote{The usage of $t_\mu$ is
straightforward (and adopted for clarity of notation) and no confusion
between the Lorentz index and the thermal index should arise.} of
the inserted photon at the vertex $\mu$. Notice that all the thermal
powers of $(-1)^{t_i+1}$ match and there is no sign ambiguity between
the relative contributions of the two terms, which is independent of
the thermal field type. Hence the two can be combined to give the total
contribution to Fig.~\ref{fig:B4} as a difference of two terms, viz.,
\begin{align}
{\cal M}_{n+1}^{q\mu,{\rm tot}} & = e^{s+1}
	(-1)^{\hbox{(}\sum_{i=1}^s t_i\hbox{)}+s}
	\cdots \left[ S\strut^{q-1,q}_{p'+\sum_{q-1}}
	\delta_{t_\mu,t_q} \left(2p'+
	2\Sigma_{q-1}+l_q\right)_{\mu_q} \right. \nonumber \\
 & \hspace{3cm} \left. - S\strut^{q-1,q}_{p'+\sum_{q-1}+k}
 	\delta_{t_\mu,t_{q-1}}
	\left(2p'+2\Sigma_{q-1}+2k+l_q\right)_{\mu_q} \right]
	S\strut^{q,q+1}_{p'+\sum_q+k} \cdots ~.
\label{eq:circledq}
\end{align}
This is the thermal generalisation of the corresponding result obtained by
GY for the fermionic case at $T=0$. This combination of {\em differences}
of terms from $K$ photon insertion helps in pair-wise cancellation and
hence simplification and factorisation of the IR divergent part even at
finite temperature. Note that due to the absence of 4-point vertices,
the corresponding thermal result for the insertion of a thermal virtual
$K$ photon into a {\em fermion} line was much simpler \cite{Indu}:
\begin{align}
{\cal M}_{n+1}^{q\mu,{\rm fermion}} & = e^{s+1}
	(-1)^{\hbox{(}\sum_{i=1}^s t_i\hbox{)}+s}
	\left[ S\strut^{q-1,q}_{p'+\sum_{q-1}}
	\delta_{t_\mu,t_q} -
	S\strut^{q-1,q}_{p'+\sum_{q-1}+k}
	\delta_{t_\mu,t_{q-1}} \right]~ 
	S\strut^{q,q+1}_{p'+\sum_q+k} \cdots ~,
\label{eq:fermion}
\end{align}
where the propagators are now fermionic. We now apply this
simplification to all sets of diagrams. We have the
following possibilities:
\begin{enumerate}
\item The inserted $K$ photon vertices are on different external lines,
in-coming and out-going.

\item The two vertices of the inserted $K$ photon are on the same lines.

\end{enumerate}
We will address them one by one.

\subsubsection{$K$ photon insertions on different lines}

\begin{figure}[bht]
\includegraphics[width=\textwidth]{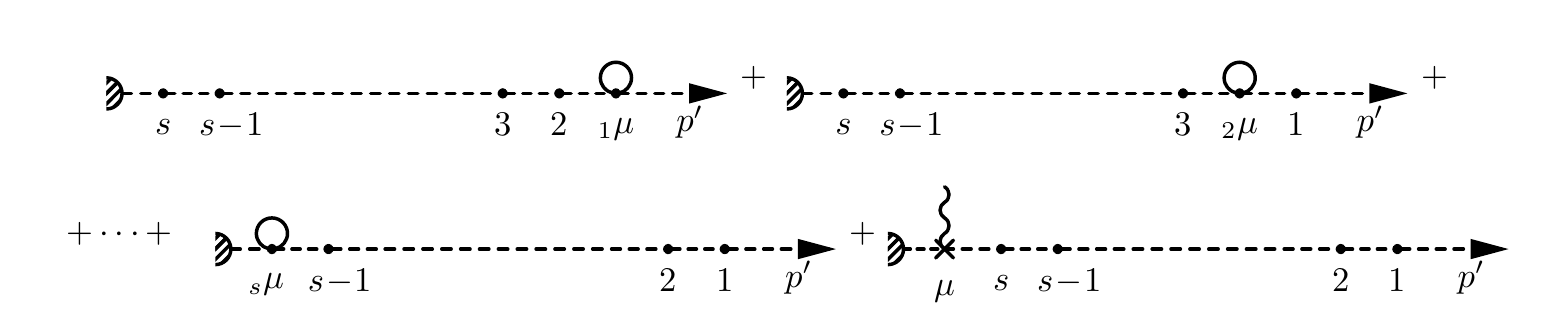}
% \caption*{Fig.~B.5}
\caption{\em The graphs in Figs.~\ref{fig:B2} and \ref{fig:B3} can be
combined into the $s$ circled vertex graphs and an $(s+1)^{\rm th}$
graph with the inserted $\mu$ vertex to the left of all the other $s$
vertices on the $p'$ leg, as shown above.}
\label{fig:B5} % actually moved to appendix in T=0, so Fig 9
\end{figure}

The case where the vertices are on different lines is straightforward.
Start with a lower order diagram that contains only 3-point vertices;
we will relax this condition later. Consider the insertion of the $\mu$
vertex of the $(n+1)^{\rm th}$ $K$ photon in all possible ways
on the $p'$ leg. In terms of the circled vertices, these can be
expressed in terms of the graphs shown in Fig.~\ref{fig:B5}.

Since the relevant term in the $K$ photon propagator is $(b_k k_\mu
k_\nu)$, we compute the contribution to the part of the matrix element,
${\cal{M}}_{n+1}^{\mu,p'{\rm leg}}$, from an insertion $\mu$ on
the $p'$ leg. The contribution from
each of the first $s$ graphs in Fig.~\ref{fig:B5}, retaining only the
$k_\mu$ term in the photon propagator, and omitting overall constants
including a factor of $e^{s+1} (-1)^{\hbox{(}\sum_{i=1}^s t_i\hbox{)}+s}$,
can be written from inspecting the result in Eq.~\ref{eq:circledq}
(see the corresponding graphs in Fig.~\ref{fig:B4}),
\begin{eqnarray} \nonumber
{\cal{M}}_{n+1}^{\mu,p',s} & \propto & \left\{0 + \delta_{t_\mu,t_1}
 	(2p'+l_1)_{\mu_1} S\strut^{t_1,t_2}_{p'+\sum_1+k}
	(2p'+2\Sigma_1+2k+l_2)_{\mu_2} \cdots (V)
	\cdots \right\} \nonumber \\ % 1mu
& & + \left\{ (2p'+l_1)_{\mu_1}
	\left[S\strut^{t_1,t_2}_{p'+\sum_1}
	\delta_{t_\mu,t_2} (2p'+2\Sigma_1+l_2)_{\mu_2} 
	\right. \right. \nonumber \\
 & & \left. \left. -S\strut^{t_1,t_2}_{p'+\sum_1+k}
 	\delta_{t_\mu,t_1} (2p'+2\Sigma_1+2k+l_2)_{\mu_2}
	\right] \cdots (V) \cdots \right\} \nonumber \\ %2mu
& & + \left\{\strut \cdots \right\} \nonumber \\
& & + \left\{ (2p'+l_1)_{\mu_1}
	S\strut^{t_1,t_2}_{p'+\sum_1} \cdots
        \left[S\strut^{t_{s-1},t_s}_{p'+\sum_{s-1}}
 	\delta_{t_\mu,t_s} (2p'+2\Sigma_{s-1}+l_s)_{\mu_s}
	\right. \right. \nonumber \\
 & & \left. \left. - S\strut^{t_{s-1},t_s}_{p'+\sum_{s-1}+k}
 	\delta_{t_\mu,t_{s-1}} (2p'+2\Sigma_{s-1}+2k+l_s)_{\mu_s}
         \right] \cdots (V) \cdots \right\}~, \nonumber \\
 & = & \left\{\strut 0 + M_1 \right\} 
 + \left\{\strut M_2 - M_1 \right\}
 + \left\{\strut \cdots \right\}
 + \left\{\strut M_s - M_{s-1} \right\}~.
\label{eq:sT}
\end{eqnarray}
Here $(V)$ denotes the (arbitrary) vertex that separates the $p'$ and
$p$ legs, and the first term vanishes since $p'$ is on-shell.
It can be seen that the terms now cancel, just as happened in the
$T=0$ case for GY, leaving only the last term, $M_s$. The contribution from
the unpaired $(s+1)^{\rm th}$ term which is the last graph shown in
Fig.~\ref{fig:B2} is
\begin{align} \nonumber
{\cal{M}}_{n+1}^{\mu,p',s+1} & \propto 
	(2p'+l_1)_{\mu_1} S\strut^{t_1,t_2}_{p'+\sum_1} \cdots
        \left[S\strut^{t_s,t_V}_{p'+\sum_s} \delta_{t_\mu,t_V} 
         - S\strut^{t_s,t_V}_{p'+\sum_s+k} \delta_{t_\mu,t_s} 
	\right] \, (V) \cdots~, \nonumber \\
	& = \left\{\strut M_{s+1} - M_s\right\}~.
\label{eq:splus1T}
\end{align}
Hence the second term of Eq.~\ref{eq:splus1T} cancels the contribution of
the previous $s$ terms in Eq.~\ref{eq:sT}, so that the total contribution
from the insertion of the $\mu$ vertex of the $(n+1)^{\rm th}$ $K$
photon in all possible ways on the $p'$ leg gives a contribution that is
independent of the inserted momentum, $k$, as in the case with fermions.
That is, the result of adding the contributions of inserting both A and B
types of vertices in all possible ways on the $p'$ leg is,
\begin{align}
{\cal{M}}_{n+1}^{\mu,p',tot} & \propto 
	(2p'+l_1)_{\mu_1} S\strut^{t_1,t_2}_{p'+\sum_1} \cdots
        S\strut^{t_s,t_V}_{p'+\sum_s} \left[\delta_{t_\mu,t_V} 
	\right] \, (V) \cdots~.
\label{eq:MnKp'T}
\end{align}
Note the presence of the delta-function, $\delta_{t_\mu,t_V}$, arising
from matching the field types at the vertex. Since the hard photon is
observable, $t_V=1$ and hence $t_\mu=1$ as well.

\paragraph{Inclusion of the 4-point vertex}

The calculation can be extended to the case when there are both 3- and
4-point vertices in the $n$-photon graph. Graphs with the same number
of photons rather than the same number of vertices are grouped together,
so that the overall charge factors (powers of $\alpha$) are the same for
the entire set of diagrams. Hence the corresponding $n$-photon graph may
have fewer than $n$ vertices, and in fact will have $(m/2 + (n-m))$
vertices if $m$ of the $n$ photons participate in a 4-point vertex.
For such diagrams there is an additional constraint since it is obvious
that the additional $(n+1)^{\rm th}$ photon cannot be added at an
already existing 4-point vertex.

Two photons, say $l_q$ and $l_r$, are at vertex $q$. No
more photons can be added at this vertex, and in fact, the vertex
factor for this vertex is proportional to $g_{qr}\delta_{t_q,t_r}$,
with no momentum dependence. As before, any $q=\mu$ vertex (that is,
the new photon forms a 4-point vertex) contributes a term with a factor
$(-2k_{\mu_q})$ in the numerator which cancels a similar term from a
3-point $\mu$ vertex as shown in Fig.~\ref{fig:B4}. The terms cancel
diagram by diagram, similar to that shown in Eq.~\ref{eq:sT}. The $g_{qr}
\delta_{t_q,t_r}$ factor gets carried along and does not spoil the
re-grouping and cancelling of terms when an additional $(n+1)^{\rm th}$
$K$-photon vertex $\mu$ is added.

A similar result is obtained when the $\nu$ vertex of the virtual
$K$ photon is inserted on the $p$ (distinct) leg, with pair-wise
cancellations, leaving a single term containing $\delta_{t_\nu,t_V}$.
Putting back the factors of $b_k(p',p)$ as well as the rest of the
photon propagator, the total
contribution from the insertion in all possible ways of an $(n+1)^{\rm
th}$ $K$-photon (contributing a factor $(b_k(p',p) k_\mu k_\nu)$) into
a set of graphs with $n$ photons containing an arbitrary number of 3-
or 4-point vertices, is given by,
\begin{align}
{\cal M}\strut^{p'p,K\gamma}_{n+1} & = -ie^2 \int \frac{{\rm d}^4
k}{(2\pi)^4} \, \delta_{t_\mu,t_V} \, \delta_{t_\nu,t_V} b_k(p',p)
D^{t_\mu,t_\nu} (k) \times {\cal M}_n~,
\label{eq:Kgp'p}
\end{align}
and hence is proportional to the lower order matrix element ${\cal
M}_n$.

The major difference between the $T=0$ case and the thermal case is the
presence of the thermal indices. Crucially, there are additional delta
functions, $\delta_{t_\mu,t_V}$ and $\delta_{t_\nu,t_V}$, arising from
matching the field types at the special (hard) scalar-photon vertex
$V$. Since the hard photon is observable, $t_V=1$ so $t_\mu$, $t_\nu=1$
as well; hence the $K$ photon thermal propagator is constrained to be of
type $D^{11}$ alone. This is a crucial requirement for the cancellation
to occur between real and virtual photon contributions to the lower
order diagram, as we shall see below.

\subsubsection{Both $K$ photon insertions on the $p'$ leg alone}
% \label{sssec:vkp'p'}

The case where both vertices of the $(n+1)^{\rm th}$ $K$ photon are
inserted on the $p'$ leg is more complex due to the presence of a large
number and type of diagrams as well as the presence of the additional 
4-point vertices. While only seagull 4-point vertices contribute in the
previous case (of independent insertions of $\mu$ and $\nu$ on different
legs), tadpole diagrams also contribute when both vertices are inserted at
the same point on the same leg. Double-counting is avoided by insisting
that the $\mu$ vertex is always to the right of the $\nu$ vertex.

\begin{figure}[htp]
\includegraphics[width=\textwidth]{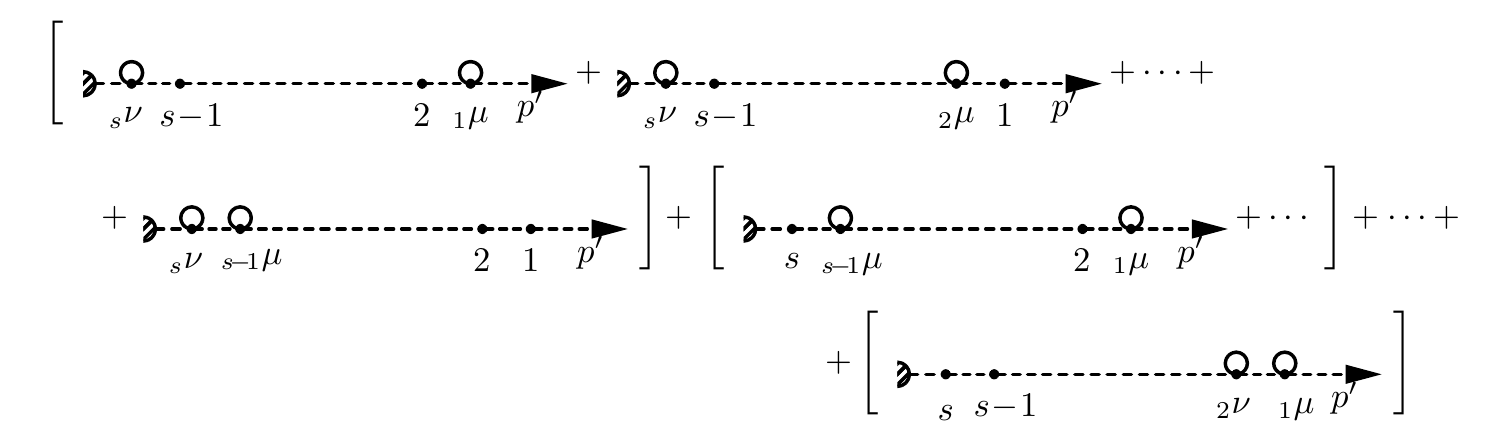}
% \caption{\em Fig.~B.6}
\caption{\em The diagrams with all circled vertices that belong to Set I.}
\label{fig:B6}
\end{figure}

\begin{figure}[htp]
\includegraphics[width=\textwidth]{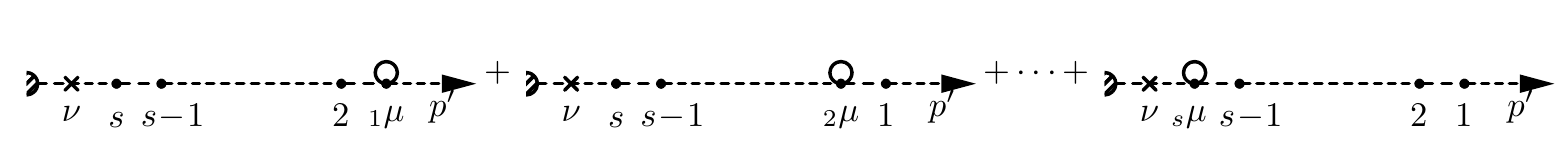}
% \caption*{Fig.~B.7}
\caption{\em The diagrams with only $\mu$ vertices circled that belong to
Set II.}
\label{fig:B7}
\end{figure}

\begin{figure}[tbh]
\begin{center}
\includegraphics[width=1.0\textwidth]{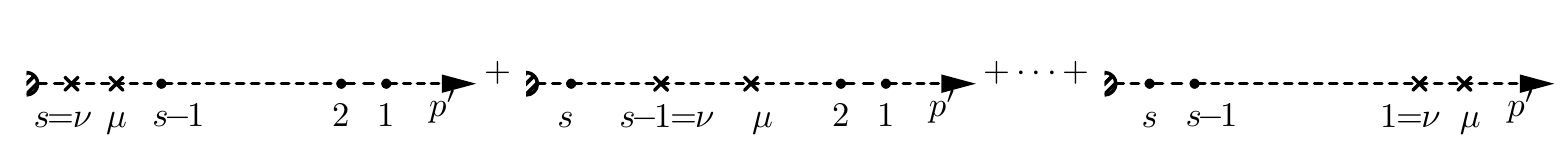}
\end{center}
% \caption*{Fig.~B.8new}
\caption{\em The diagrams with only $q=\nu$ vertices circled that belong to
Set III.}
\label{fig:B8new}
\end{figure}

\begin{figure}[tbh]
\begin{center}
\includegraphics[width=1.0\textwidth]{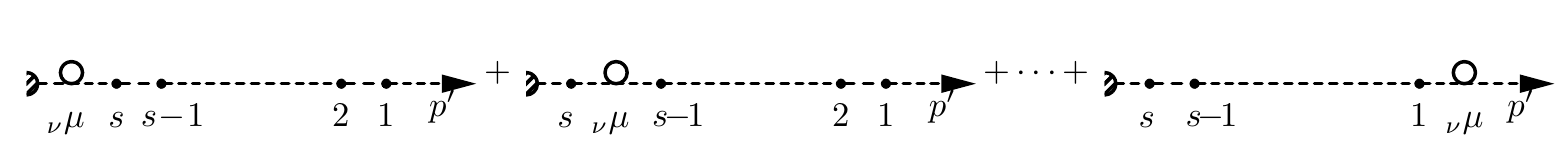}
\end{center}
% \caption*{Fig.~B.9new}
\caption{\em The diagrams with ${}_\nu\mu$ circled vertices that belong to
Set IV. The last term corresponds to the self energy
diagram and is to be omitted on the $p$ leg to avoid double counting.}
\label{fig:B9new}
\end{figure}

\begin{figure}[htp]
\begin{center}
\includegraphics[width=0.5\textwidth]{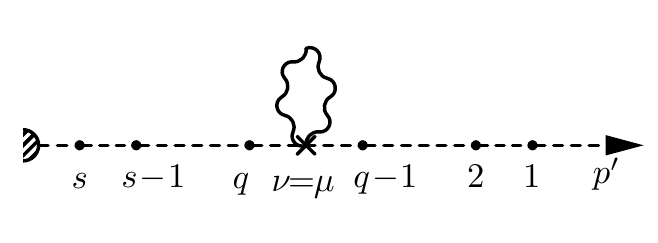}
% \caption*{Fig.~B.10}
\caption{\em A typical tadpole diagram where the vertices $\mu$, $\nu$ of
the $(n+1)^{\rm th}$ $K$ photon have been inserted between the vertices
$q$ and $q-1$.}
\label{fig:B10}
\end{center}
\end{figure}

As before, the case with only 3-point vertices in the lower order
graph is first considered; this condition is then relaxed to prove
the general case. The diagrams obtained when the $(n+1)^{\rm th}$
$K$ photon is inserted in all possible ways can be grouped into
sets labelled Set I, Set II, Set III, and Set IV, as shown in
Figs.~\ref{fig:B6} to \ref{fig:B9new}. While Set I (Fig.~\ref{fig:B6})
has circled vertices at both $\mu$ and $\nu$ insertions, Set II
(Fig.~\ref{fig:B7}) has circled vertices only at $\mu$, with $\nu$ to the
right of the special $V$ vertex. Set III (Fig.~\ref{fig:B8new}) has all
4-point vertex insertions at $\nu$, with $\mu$ immediately adjacent to
$\nu$. Finally, Set IV (Fig.~\ref{fig:B9new}) is a set of ${}_\nu\mu$
circled vertices that includes all tadpole insertions, $\mu=\nu$, as
shown in Fig.~\ref{fig:B10}.

In order to simplify the analysis, we first consider only the $T=0$
contribution which is logarithmically divergent in the IR/soft limit. We will
show that various contributions can be combined so that the term-by-term
cancellation is more easily seen. This analysis also highlights the role
of the tadpole contributions in factorising the IR divergent parts. We
will then use the understanding acquired in this analysis to consider
the entire finite temperature contribution which contains linear
divergences and logarithmic subdivergences that must be factorised as well.

\paragraph{$K$ photon insertions with both vertices on the $p'$ leg:
$T=0$}

Tadpole diagrams are proportional to $g_{\mu\nu}$ as per the Feynman
rules in Appendix \ref{app:Ascalar}. (Note the presence of an additional
symmetry factor of $1/2$ with respect to the seagull vertex factor shown
here.) Hence the $K$ photon insertions of
tadpoles on the $p'$ leg contribute terms proportional to $k^2$ and
are IR finite. Thus, it appears at first sight that tadpole diagrams can
be neglected when discussing the IR behaviour of scalar QED. However,
this is not so, since the contributions from these tadpoles are crucial
in obtaining the factorisation and subsequent exponentiation of the IR
divergent terms from the $K$ photon insertions. We show this result by
first considering the (simpler) $T=0$ case without tadpoles. The
contributing diagrams are Sets I to IV, excluding the tadpole
contributions in Set IV, which we name as Set IV$'$.

In Appendix ~\ref{app:vkgamma} we show that the contributions from
various contributing diagrams from Sets I to IV$'$ can be expressed as a
term proportional to the lower order matrix element, ${\cal{M}}_n$, as
required, as well as terms linear and quadratic in $k$:
\begin{eqnarray} \nonumber
{\mathcal{M}}_{n+1}^{\mu\nu,p'p'} (T=0) & \propto &
  \left[ \left\{ (2p'+l_1)_{\mu_1} \frac{1}{(p'+\mathcal{S}_1)^2}
	 \cdots \frac{1}{(p'+\mathcal{S}_s)^2} (V) 
	 \cdots \right\} \right] \nonumber \\
 & & -\left[ \left\{(2p'+l_1)_{\mu_1} \frac{1}{(p'+\mathcal{S}_1)^2}
	 \cdots \frac{1}{(p'+\mathcal{S}_{s-1})^2} (-2k)_{\mu_s}
        \frac{1}{(p'+\mathcal{S}_s)^2} (V) \cdots \right\} \right.
 + \left\{  \cdots \right\}  \nonumber \\
  & & + \left\{(2p'+l_1)_{\mu_1} \frac{1}{(p'+\mathcal{S}_1)^2}
  	\cdots {(2p'+2\Sigma_{q-1} + l_q)_{\mu_q}}
        \frac{1}{(p'+\mathcal{S}_q)^2} (-2 k)_{\mu_q} \cdots
	(V) \cdots \right\} \nonumber \\
 & & + \left. \left\{  \rule{0mm}{5mm} \cdots \right\} \right]
	 +\left[ \left\{  \rule{0mm}{5mm} \cdots
 \right\} \right.  + \left\{(2p'+l_1)_{\mu_1} \frac{1}{(p'+\mathcal{S}_1)^2}
  \cdots (2p'+2\Sigma_{q-1}+l_q)_{\mu_q} \times \right.
  \nonumber \\
 & & ~~~\left\{ \frac{1}{(p'+\mathcal{S}_q)^2} {(2 k \cdot (p'+
	\sum_q) + k^2)}
        \frac{1}{(p'+\mathcal{S}_q)^2} \right\}
        (2p'+2{\sum}_q+l_{q+1})_{\mu_{q+1}} \cdots \nonumber \\
 & &   ~~~\left. \frac{1}{(p'+\mathcal{S}_s)^2} (V) \cdots \right\} 
  	+ \left. \left\{ \rule{0mm}{6mm} \cdots \right\} \right]~.
% \end{multlined}
\label{eq:Kp'p'}
% \end{align}
\end{eqnarray}
Here the zero temperature propagator of a scalar with momentum
$p'+\sum_q$ is $i/((p'+\sum_q)^2-m^2)$,
which we have represented as $i/(p'+{\cal{S}}_q)^2$. It can be seen
that apart from the first term in Eq.~\ref{eq:Kp'p'} which is proportional to
the lower order matrix element, ${\cal{M}}_n$, the remaining terms are
proportional to $(-2k)_{\mu_q}$, $2k \cdot P$, and $k^2$, {\em with no
other $k$ dependence in the denominators}. Since $b_k$ is even in $k$
(by definition), the terms proportional to either of $(-2k)_{\mu_q}$ and
$2 k \cdot P$ vanish on integration, leaving only the term proportional
to ${\cal{M}}_n$ and terms proportional to $k^2$.  Recall that the $T=0$
contribution is logarithmically divergent in the soft photon limit
while the finite $T$ contribution has both linear (${\cal{O}}(1)$) and
logarithmic (${\cal{O}}(k)$) divergences; consequently, terms proportional
to $k^2$ are IR finite. While these ${\cal{O}}(k^2)$ terms do not spoil
the IR finiteness of the theory, they also do not allow the $(n+1)^{\rm
th}$ matrix element to be expressed as purely proportional to the
lower $n^{\rm th}$ order matrix element, which would have enabled the
IR divergent pieces to be factorised and exponentiated to all orders,
eventually cancelling with the corresponding IR divergent parts from
the real photon contributions.

A straightforward calculation of the contribution of the hitherto
neglected tadpole diagrams in Set IV (see Appendix \ref{app:vkgamma}
shows that this contribution exactly cancels these ${\cal{O}}(k^2)$ terms
left over in Eq.~\ref{eq:Kp'p'} above; hence the total contribution
from inserting a vritual $K$ photon in all possible ways such that
both vertices are on the $p'$ leg is simply a term proportional to the
lower order matrix element, as required. With the understanding that
the tadpole contributions are crucial in achieving this result, we now
go on to consider the finite temperature case of interest.

\paragraph{$K$ photon insertions with both vertices on the $p'$ leg:
finite $T$}

We use the insight we have gained from the zero temperature case to
complete the calculation for the general thermal case when both the
$K$ photon vertices are inserted into the same leg. The same diagrams
contribute; we consider each set in turn and make use of the generalised
thermal identities discussed in Appendix \ref{app:fidentities} since
the propagators are no longer simple and have a $2 \times 2$ matrix
form. Generalised identities are used to get term by term
cancellations; as in the zero temperature case, there are many left-over
terms due to the presence of the additional 4-point vertices, in contrast
to the fermionic case. This is the most complex of the calculations, and
most of the details are relegated to Appendix \ref{app:vkgamma}. The
grouping of terms in order to cancel them is made easier by the
understanding gained from considering the zero temperature case. 

As in the zero temperature case, there are many terms contributing to
the various Sets. It is shown in Appendix~\ref{app:vkgamma} that
left-over terms of Sets $(I+II+III)$ cancel against corresponding
terms in Set IV, leaving behind a term proportional to the lower order
matrix element ${\cal M}_n$, as required, and towers of terms linear
in $k$. The cancellation occurs as follows. The contribution from Set IV
can be expressed as,
\begin{align}
{\cal{M}}_{n+1}^{\mu\nu,p'p', IV} & \propto
 \left\{ \left[A_s - B_s + C_s \right] \right\} +
 	\left\{\cdots\right\} +
  	\left\{ \left[ A_1 - B_1 + C_1 \right] \right\} + 
  	\left\{ \left[ -B_0 + C_0 \right] \right\}~,
\label{eq:Mp'p'4main}
\end{align}
where 
\begin{align}
A_q & = (2p'+l_1)_{\mu_1}
        S\strut^{12}_{p'+\sum_1}
	(2p'+2\Sigma_1+l_2)_{\mu_2}
	\cdots \left[S\strut^{q-1,q}_{p'+\sum_{q-1}+k}
	\delta_{t_\mu,t_{q-1}} \delta_{t_\nu,t_q} \right] \cdots
	(\hbox{no } k) ~, \nonumber \\
B_q & = (2p'+l_1)_{\mu_1}
        S\strut^{12}_{p'+\sum_1}
	(2p'+2\Sigma_1+l_2)_{\mu_2}
	\cdots \left[S\strut^{q-1,q}_{p'+\sum_{q-1}}
	\delta_{t_\mu,t_{q-1}} \delta_{t_\nu,t_{q-1}} \right] \cdots
	(\hbox{no } k) ~, \nonumber \\
C_q & = (2p'+l_1)_{\mu_1}
        S\strut^{12}_{p'+\sum_1}
	(2p'+2\Sigma_1+l_2)_{\mu_2}
	\cdots \left[S\strut^{\nu,q}_{p'+\sum_{q}}
  	(2p'+2\Sigma_{q-1})\cdot k \, \delta_{t_\mu,t_\nu} 
  	S\strut^{\nu,q}_{p'+\sum_q} \right] \cdots
	(\hbox{no } k) ~,
\end{align}
where ``no $k$'' indicates that the remaining terms do not have any $k$
dependence. Here the term $(-B_0+C_0)$ arises from self-energy
corrections on the $p'$ leg, with $B_i$ proportional to ${\cal M}_n$.
Note that all $C_i$ are odd in $k$.

Details of the calculation for Sets I, II, and III are given in Appendix
\ref{app:vkgamma}. Combining Sets I, II and III, we have,
\begin{align}
{\cal{M}}_{n+1}^{\mu\nu,p'p', I+II+III} & = 
\left[\sum_{q=1}^s B_q - \sum_{q=0}^s A_q + Y\right]
% \label{eq:Mp'p'123}
\end{align}
where $Y$ is defined in Eq.~\ref{eq:Mp'p'123} in Appendix
\ref{app:vkgamma} and each term in $Y$ is proportional to
$(-2k)_{\mu_q}$. Hence both the contributions of $C_q$ and $Y$ have
linear powers of $k$ in the numerator, and {\em no other dependence
on $k$} apart from the overall factors such as $b_k$, etc. Hence these
terms are odd in $k \leftrightarrow -k$ and vanish. We see that the $A_q$
terms cancel between the contributions of Set IV in Eq.~\ref{eq:Mp'p'4}
and Sets ($I+II+III)$ in Eq.~\ref{eq:Mp'p'123}, as also all the $B_q$
except for $-B_0$ which remains as the only left over piece when all
sets are combined and we further recall that this term is proportional
to the lower order matrix element, as in the $T=0$ case.

Note that the ${\cal O} (k^2)$ terms terms are present in the two
contributing graphs corresponding to each term in the typical circled
vertices of Set IV and exactly cancel against one another. While the
${\cal O} (k^2)$ terms, and hence, the tadpole contributions are IR
finite and do not pose any problems for the theory, it is not just
simply a preference that these be included with the IR finite $G$ photon
contributions; $b_k$ was designed to isolate the IR singular terms and
resum them; hence the presence of such ${\cal O} (k^2)$ terms in addition
to the term proportional to the lower order matrix element, precludes
the factorisation and resummation of the $K$ photon contributions to
all orders; hence it is a matter of satisfaction that such ${\cal O}
(k^2)$ terms cancel exactly.

The sum of the contributions from all four sets of diagrams with all
possible insertions of the $(n+1)^{\rm th}$ $K$ photon, with both
vertices on the $p'$ leg, is therefore a term that contains the IR divergence
and is proportional to the lower order matrix element {\em with
no additional finite contributions}, as is the case with the zero
temperature theory and fermionic QED. Putting back the overall factors,
we have,
\begin{equation}
{\cal M}\strut^{p'p', K\gamma}_{n+1} = +ie^2
	\int \frac{{\rm d}^4 k}{(2\pi)^4} \, \delta_{t_\mu,t_1} \,
	\delta_{t_\nu,t_1} \, b_k(p',p') \, D^{t_\mu,t_\nu} (k)~{\cal M}_n~.
\label{eq:Kgp'p'}
\end{equation}
Since $t_1=1$ necessarily, it depends only on the $D^{11}$ photon
propagator, as before.

\subsubsection{Both $K$ photon insertions on the $p$ leg alone}

A similar analysis can be done for the case when both the vertices of
the inserted $(n+1)^{\rm th}$ $K$ photon are on the $p$ leg. As
discussed in GY, the outermost self energy insertion graph is
neglected here to compensate for wave function renormalisation, due to
which the sum of contributions for all possible insertions on the $p$
leg adds up to zero. As shown in Appendix \ref{app:vkgamma}, the term
$(-B_0 + C_0)$ arises from self-energy corrections on the $p'$ leg. A
similar contribution occurs when the virtual $K$ photon is inserted on
the $p$ leg. Hence when we remove the self-energy correction on the $p$
leg to account for wave function renormalisation, we remove the term
corresponding to $(-B_0 + C_0)$. (The contribution of $C_0$ is in any
case zero since it is odd in $k$.) The total contribution for the insertion
of a virtual $K$ photon in all possible ways on the $p'$ leg added up to
$-B_0$. When this is removed to account for wave function
renormalisation for insertions on the $p$ leg, we find that
the total contribution vanishes.

Since this compensation could have been included
in either of the legs, we symmetrise over the two possibilities, thus
giving us the contributions:
\begin{align} \nonumber
{\cal M}\strut^{p'p', K\gamma}_{n+1} & = +ie^2 \frac{1}{2}\,
	\int \frac{{\rm d}^4 k}{(2\pi)^4} \, \delta_{t_\mu,t_1} \,
	\delta_{t_\nu,t_1} \, b_k(p',p') \, D^{t_\mu,t_\nu} (k)~{\cal M}_n~,
	\nonumber \\
{\cal M}\strut^{pp, K\gamma}_{n+1} & = +ie^2 \frac{1}{2} \,
	\int \frac{{\rm d}^4 k}{(2\pi)^4} \, \delta_{t_\mu,t_1} \,
	\delta_{t_\nu,t_1} \, b_k(p,p) \, D^{t_\mu,t_\nu} (k)~{\cal M}_n~.
\label{eq:Kgpp}
\end{align}
The contribution is once more proportional to the lower order matrix
element and depends on the $D^{11}$ part of the inserted photon propagator
alone.

\subsubsection{Inclusion of `disallowed diagrams'}

Certain `disallowed diagrams' may contribute at higher orders. For
instance, the outermost self-energy insertion graph is removed at a
certain order to account for wave function renormalisation. However,
while making $K$ or $G$ photon insertions at the next higher order,
these lower order diagrams must be included, as these can give rise to
allowed graphs at the next order. As in the case of the zero
temperature theories, these terms add to zero. There is an additional
disallowed diagram in the thermal case that must be similarly included:
these are lower order graphs with `outermost' vertices next to the $p'$
or $p$ external legs that are of thermal unphysical type with $t_1=2$. A
calculation shows that these diagrams also do not contribute at the next
higher order.

\subsubsection{The total $K$ photon contribution}

The total contribution from the insertion of the $(n+1)^{\rm th}$ virtual
$K$ photon therefore is,
\begin{align} \nonumber
{\cal{M}}_{n+1}^{K\gamma,{\rm tot}} = & \frac{ie^2}{2} \int
	\frac{{\rm d}^4 k}{(2\pi)^4} \, \left\{ \delta_{t_\mu,t_1} \,
	\delta_{t_\nu,t_1} \, D^{t_\mu,t_\nu} (k) \,
	\left[\strut b_k(p',p') + b_k(p,p) \right] \right. \nonumber \\
	 & +\left. \delta_{t_\mu,t_V} \, \delta_{t_\nu,t_V} \,
	 D^{t_\mu,t_\nu} (k) \,
	\left[\strut -2b_k(p',p) \right] \right\}
	 {\cal{M}}_{n}~, \nonumber \\
	 & \equiv \left[B \right] {\cal{M}}_{n}~,
\label{eq:K}
\end{align}
where
\begin{align} \nonumber
B & = \frac{ie^2}{2} \int
	\frac{{\rm d}^4 k}{(2\pi)^4} \, D^{11} (k) \,
	\left[\strut b_k(p',p') - 2 b_k(p',p) + b_k(p,p) \right]~, 
	 \nonumber \\
  & \equiv \frac{ie^2}{2} \int
	\frac{{\rm d}^4 k}{(2\pi)^4} \,
	D^{11} (k) \, \left[\strut J^2(k) \right]~.
\label{eq:B}
\end{align}
In Eq.~\ref{eq:B} we have used the fact that the thermal types of the
hard/external vertices must be type-1; $t_1=t_V =1$, so that each term
is proportional to the (11) component of the photon contribution.
This will be crucial to achieve the cancellation between virtual and real
photon insertions, as we show below.

Hence the {\em structure} of the contribution from virtual $K$ photon
insertion is the same as in the $T=0$ case; however, note that, due to the
thermal contributions in the photon propagator, there are both linear
and logarithmic divergences in these terms. Demonstration of the
cancellation of the linear divergences follows the same route as that of
GY at $T=0$; the demonstration of cancellation of the logarithmic
subdivergences is discussed separately later.

\subsection{Insertion of virtual $G$ photons}
\label{ssec:vggamma}

In GY, it was shown that insertion of a virtual $G$ photon into the $n$
vertex graph with only 3-point vertices gives finite contributions.
The key point was that the $G$ photon contribution at $T=0$ was
proportional to
\begin{align} \nonumber
{\cal{M}}_{n+1}^{G\gamma;T=0} & \propto
	\left\{ g_{\mu\nu} - b_k(p_f,p_i) k_\mu k_\nu\right\} \times
	p_f^\mu \, p_i^\nu~, \nonumber \\
 & = 0 + {\cal{O}}(k)~.
\label{eq:Gfermion}
\end{align}
Since the leading divergence for the $T=0$ theory is a logarithmic one,
terms proportional to powers of $k$ in the numerator are IR finite;
hence the $G$ photon contribution was IR finite.

At finite temperatures, there are two major modifications: one due to
the thermal part of the photon propagator and the other
due to the thermal part of the scalar propagator. We start by
considering the contribution due to the thermal part of the photon
propagator. Although there are different types of thermal fields and
hence four different photon propagators, ${\cal D}^{ab}_{\mu\nu}$, all
of them have the same leading IR behaviour: the divergence is a linear
one due to the presence of the term in the photon propagator that is
proportional to
\begin{equation}
2\pi \delta (k^2) N(\vert k^0 \vert) \equiv 2\pi \delta(k^2)
\frac{1}{\exp^{\vert k^0 \vert/T}  - 1}~.
\end{equation}
This is cancelled for the $G$ photons in exactly the same way as the
$T=0$ case. However, there is also a logarithmic subdivergence arising
from terms linear in $k$ in the numerator whereas these terms are IR
finite in the $T=0$ case. Proving the IR finiteness of these
contributions is the central result of this paper. A detailed
case-by-case analysis can be found in Appendix \ref{app:vggamma}.

We start by ignoring the $T=0$ parts of the propagators
and concentrate on the thermal parts alone.  Since the thermal part of
the photon propagator includes an overall $\delta(k^2)$, there are two
simplifications that result. First, the coefficient factor $b_k(p_f,p_i)$
simplifies to
\begin{equation}
b^{T\ne 0}_k(p_f,p_i) = \frac{p_f \cdot p_i} {p_f \cdot k \; \; p_i \cdot k}~.
\label{eq:bkT}
\end{equation}
In addition, we can ignore $k^2$ terms in the scalar propagators.
The complete structure of this matrix element can be written as,
\begin{align}
{\cal M}\strut^{G\gamma}_{n+1} & \sim \int d^4 k 
	\left[ \frac{i}{k^2+i\epsilon} \delta_{t_\mu,t_\nu} \pm  2 \pi
	\delta(k^2) N(\vert k \vert) D^{t_\mu,t_\nu}_T \right] \,
	\left[\strut g^{\mu\nu} - b_k k^\mu k^\nu \right] \,
 	\left[\strut  {\rm scalar} \right]_{\mu\nu},
\label{eq:MGstr}
\end{align}
where the terms in the first two square brackets correspond to terms
in the definition of the $G$ photon propagator, with the relative
sign in the first being determined by the thermal field indices, $t_\mu,
t_\nu$. The last term represents the contribution from the $\mu$ and
$\nu$ virtual $G$ photon insertions
on the scalar legs, $p$ and $p'$, and are products of the vertex and
propagator factors. Combining the second term in Eq.~\ref{eq:MGstr}
with the vertex factors at {\em only} the $\mu$ and $\nu$ vertices
(assuming them to be 3-point for now) in the third term, we get,

\begin{align} \nonumber
 & \begin{multlined}
 \left[g^{\mu\nu} - b_k k^\mu
	k^\nu \right] \,
	\left\{S\strut^{q-1,\mu}_{p_f+\sum_{q-1}}
	\left[(2p_f+ 2\Sigma_{q-1}+k)_\mu  \,
	S\strut^{\mu,q}_{p_f+\sum_{q-1}+k}
	(2p_f+ 2\Sigma_{q-1}+2k+l_q)_{\mu_q}\right]
	\right\} \times \\
 \left\{S\strut^{m+1,\nu}_{p_i+\sum_{m}+k}
 	\left[(2p_i+ 2\Sigma_{m}+k)_\nu  \,
	S\strut^{\nu,m}_{p_f+\sum_{m}}
 	(2p_i+ 2\Sigma_{m-1}+l_m)_{\mu_m}\right]\right\}   \,
\end{multlined} \\
% \begin{align} \nonumber
 & = \left[\strut 4 P_f \cdot P_i + 2 (P_f+P_i) \cdot k - 4 b_k 
  P_f \cdot k \,P_i \cdot k \right] \left[S\strut^{q-1,\mu}
	S\strut^{\mu,q} S\strut^{m+1,\nu} S\strut^{\nu,m} (\cdots)_{\mu_q}
	(\cdots)_{\mu_m} \right]~,\nonumber \\
 & = \left[\strut 0 (p_f \cdot p_i) + 2 (p_f  + 2 p_i) \cdot k 
 	\right] \cdots ~,
\label{eq:Gpart}
\end{align}
% \end{multline}
where we have used $P_f = p_f + \sum_{i=1}^{q-1}l_i$ and
$P_i = p_i + \sum_{i=1}^{m-1}l_i$.  In the soft limit,
replacing $P_f \to p_f$, $P_i \to p_i$, and substituting for $b_k$
from Eq.~\ref{eq:bkT}, we get the last line of Eq.~\ref{eq:Gpart}. We
see that the leading $(p_f \cdot p_i)$ term vanishes (indeed, $b_k$
was chosen for this very reason) and the term in the square brackets is
exact with no further corrections. The ellipses refer to the contribution
from the remaining vertices and propagators, some of which (the set of
vertices and propagators that lie between the $\mu$ and $\nu$ vertices)
also depend on $k$. Substituting this back in Eq.~\ref{eq:MGstr}, we have,
\begin{align}
{\cal M}\strut^{G\gamma}_{n+1} \sim & \int d^4 k
	\left[ \frac{i}{k^2+i\epsilon} \delta_{t_\mu,t_\nu} \pm  2 \pi
	\delta(k^2) N(\vert k \vert) D_{t_\mu,t_\nu} \right]
        \left[\strut 0 (p_f \cdot p_i) + 2 (p_f  + 2 p_i) \cdot k \right]
 	\left[\strut {\rm scalar} \right]_{\slashed{\mu}\slashed{\nu}}~,
\label{eq:MGstrsimp}
\end{align}
where the slashes on $\mu$ and $\nu$ indicate that the contribution from
these vertices have been removed and simplified as per
Eq.~\ref{eq:Gpart} and,
\begin{align}
 \left[ {\rm scalar} \right]_{\slashed{\mu}\slashed{\nu}} & \sim 
 \left[ {\cal O}(1) + {\cal O}(k) + {\cal O}(k^2) + \cdots \right]~,
\label{eq:sca}
\end{align}
where we have indicated the powers of $k$ in the numerator of the matrix
element from the scalar contribution above.

We know that the $T=0$ part is logarithmically divergent while the
leading thermal divergence is linear. The factor $b_k$ is so chosen
so that the $(p_f \cdot p_i) \times [{\cal O}(1)]$ term, obtained by
combining Eqs.~\ref{eq:MGstrsimp} and \ref{eq:sca}, vanishes. Note that
this term gives rise to the leading log divergence at $T=0$ (from the
$1/(k^2+ i\epsilon)$ term in the photon propagator) as well as the leading
linear divergence at $T\ne 0$ (from the $\delta(k^2) N(\vert k \vert)$
term). The remaining $T=0$ part is IR finite since any power of $k$
in the numerator renders the term finite.

At $T \ne 0$, in addition, the logarithmic subdivergence arising from the
$(p_f \cdot p_i) \times [{\cal O}(k)]$ term from Eqs.~\ref{eq:MGstrsimp}
and \ref{eq:sca}, also vanishes since the coefficient of this term
is zero. But there is a term arising from the $((p_f + p_i) \cdot k)
\times [{\cal O}(1)]$ factor in the thermal part, that appears to
be a logarithmic subdivergence. We however observe that the $[{\cal
O}(1)]$ terms in the scalar part are symmetric under the interchange $(k
\leftrightarrow -k)$; since the term $((p_f + p_i) \cdot k)$ is linear in $k$,
the entire contribution is odd under this interchange, so that this
potential subdivergent log contribution vanishes. Higher order terms
arising from even powers of $k$ in the integrand are IR finite. Hence
the $G$ photon insertions are IR finite.

We have implicitly assumed that there are no divergences associated
with the photon momenta $l_i$ in the lower order graphs (that is,
from ${\cal M}_n$). This is not necessarily true; divergences can
potentially arise from any of the soft photons in the graph. Here,
the procedure, as shown by GY, is to separate out the photon momenta
into groups that cause an IR divergence and those that do not. It
is then possible to ignore the latter group and construct so-called
``skeletal graphs" where the divergence arises only when each of the
controlling momenta, $l_i, i=1,\cdots,m$, simultaneously vanish. It
was shown in Refs.~\cite{GY,Indu} that $G$ photon insertions are
finite with respect to all such controlling momenta for a theory of
charged fermions at zero and finite temperature. In
Appendix~\ref{app:vggamma} we show that this holds for scalars at finite
temperature as well.

This result also holds when we extend the analysis to include the
possibility that the $\mu$ and $\nu$ vertex insertions are of 4-point
type, or even that some or all of the vertices in the lower order
graph are of 4-point type as well; each of these cases is dealt with in
detail in Appendix~\ref{app:vggamma}. The final generalisation is when
we include thermal effects in the scalar propagator as well (those in
the vertices are quite trivial to deal with). We discuss this below.

\subsubsection{Effect of including thermal scalars}
\label{ssec:thermal}

When the scalar field is also thermal, it is not sufficient to consider
the $1/(P^2 -m^2)$ part of the scalar propagator. There are factors of
the scalar number operator, $N_S$, that can cause a potential divergence
since the scalar fields are bosons with,
\begin{equation}
N_S(\vert P^0 \vert)  = \frac{1}
	{\exp[\vert P^0 \vert /T] -1} \; 
  \stackrel{P^0 \to 0}{\longrightarrow} \; \frac{1} {\vert P^0 \vert}~,
\end{equation}
in contrast to the fermionic case where the number operator is finite,
$N_f \to 1/2$, as $\vert P^0 \vert \to 0$; so we need to check
that this result holds when the scalars are thermal as well. We begin
as usual by considering graphs with only 3-point vertices.

The numerator factors arising from the scalar-photon vertices acquire
only irrelevant modifications when temperature effects are included;
hence the structure of the vertices, that were crucial in obtaining
the cancellation of the leading divergence of the $G$ photon
contributions between the $g_{\mu\nu}$ and $b_k k_\mu k_\nu$ terms in
Eq.~\ref{eq:Gpart}, still holds. We need to consider only the terms
linear in $k$ that can give rise to subleading logarithmic divergences
as discussed above.

We, therefore, examine the finite temperature dependence of the scalar
propagators. In contrast to the case of thermal photons, the momentum
$p_f$ (or $p_i$) flows through all the scalar lines and this controls
the behaviour in the soft limit. The pure $L_0(l_i) \sim 1/(l_i^2-m^2)$
dependence at $T=0$ is replaced by a sum of $L_0(l_i)$ and $L_T(l_i)
\sim \delta (l_i^2 - m^2)$ terms. Hence, none, some, or all the scalar
propagators can have thermal contributions. The case where all scalar
propagators correspond to $L_0$ is the case that we have studied so
far.

While the two propagators have the same dimensional dependence on $k$,
$L_T(l_i = P+k)$ contains a delta-function $\delta((P+k)^2-m^2)$
dependence which either makes the term finite or else leads to a
constraint where $k^0$ is related to combinations of the remaining
(controlling) momenta and hence there is no (logarithmic sub)divergence
associated with this term. This holds even when more than one of the
scalar propagators is a thermal $L_T$ type. The detailed analysis for
adding a $G$ photon to a lower order graph with thermal electrons, and
having one or more momenta in the controlling set, can be found in
Ref.~\cite{Indu} and applies to the case of charged scalars as well.
Hence the $G$ photon insertion is IR finite when we consider the entire
thermal structure of the theory, both for charged scalars and photons,
and even if the charged particles are fermions. More details are found
in Appendix~\ref{app:vggamma}. Finally, the cases when some of the
vertices are $K$ photons or real photon vertices is also discussed
in Appendix~\ref{app:vggamma}.

As before, we have to verify that when we ``flesh out'' skeletal graphs
and include self-energy or other terms, the graph remains IR finite; this
is also shown in Appendix~\ref{app:vggamma}. This concludes the proof that
the entire virtual $G$ photon insertions of the full finite temperature
theory (with both charged fermions and scalars) are in general IR finite.

\subsection{The final matrix element for virtual photons}
\label{sec:vgamma}

We have obtained the familiar result that the (IR divergent)
contribution of the $K$ photon insertions is proportional to the lower
order matrix element, ${\cal M}_n$ while the $G$ photon insertions are
finite. We proceed as in the case of $T=0$ scalar QED or thermal
fermionic QED and consider the contribution of the $n^{\rm th}$ order
graph with $n_K$ virtual $K$ photons and $n_G$ virtual $G$
photons. Hence $n=n_K+n_G$ and there are at most $n$ vertices (since
some can be seagulls or tadpoles). As a consequence of the
Bose symmetry for the $n$ photons, each
distinct graph can arise in $n!/n_K!\,n_G!$ ways, so that the total
matrix element can be expressed as a sum of all possible individual
contributions,
\begin{equation} \frac{1}{n!}\, {\cal M}_n =
  \sum_{n_K=0}^{n} \frac{1}{n_K!}  \frac{1}{n - n_K!} {\cal
    M}_{{n_G},{n_K}}~.
\end{equation}
Summing over all orders, we get
\begin{align}
\sum_{n=0}^{\infty} \frac{1}{n!} {\cal M}_n & =
	\sum_{n=0}^{\infty}
	\sum_{n_K=0}^{n} \frac{1}{n_K!}
	\frac{1}{n - n_K!} {\cal M}_{{n_G},{n_K}}~, \nonumber \\
 & = \sum_{n_K=0}^\infty
 	\sum_{n_G=0}^{\infty} \frac{1}{n_K!}
	\frac{1}{n_G!} {\cal M}_{{n_G},{n_K}}~,
\end{align}
and we use the result that the $K$ photon contribution is proportional
to the lower order matrix element to obtain:
\begin{align} {\cal M}_{{n_G},{n_K}} &
= (B)^{n_K} M_{n_G,0} \equiv (B)^{n_K} M_{n_G}~,
\end{align}
where $B$ as defined in Eq.~\ref{eq:B} is the contribution from each
$K$-photon insertion and can be isolated and factored out, leaving only
the IR finite $G$-photon contribution, ${\cal M}_{n_G}$. Re-sorting and
collecting terms, we obtain the requisite exponential IR divergent factor:
\begin{align}
\sum_{n=0}^\infty \frac{1}{n!}
{\cal M}_n & = \sum_{n_K=0}^\infty \frac{(B)^{n_K}}{n_K!}
	\sum_{n_G=0}^{\infty}
	\frac{1}{n_G!} {\cal M}_{n_G}~, \nonumber \\
& = {\rm e}^{B} \sum_{n_G=0}^\infty \frac{1}{n_G!}
	{\cal M}_{n_G}~.
\label{eq:Mvirtual}
\end{align}
Again we highlight that this factorisation was made possible since the
$K$-photon insertions gave precisely one term and no additional pieces,
IR-finite or otherwise; this occurred due to the presence of both 3-point
and 4-point vertices in the theory. The resulting cross section including
only the virtual photon contributions to all orders is,
\begin{align} \nonumber
\sigma^{\rm virtual} & \propto \int \d\phi_{p'}
	(2\pi)^4 \delta^4(p+q-p') \left\vert
\sum_{n=0}^\infty \frac{1}{n!} {\cal M}_n \right\vert^2~,
\nonumber \\
 & = \int \d\phi_{p'} (2\pi)^4 \delta^4(p+q-p') ~\vert Z \vert^2 ~
 \sigma^{\rm virtual}_G~,
\label{eq:sigmav}
\end{align}
where $\d\phi_{p'}$ is the phase space factor corresponding to the final
state scalar with momentum $p'$ and a(n irrelevant) flux factor in the
denominator has been suppressed. The IR-finite part is contained in the
last term and the IR divergent part is contained in the exponent,
\begin{align}
\vert Z \vert^2 & \equiv \exp \left(\strut B + B^* \right)~,
\label{eq:Z2}
\end{align}
and will be shown below to cancel against a corresponding contribution
from real (soft) photon emission/absorption with respect to the heat
bath, thus indicating that thermal scalar electrodynamics is also IR
finite at all orders.

\subsection{Emission/absorption of real photons}
\label{ssec:rgamma}

There is a major difference in the thermal case: real photons can be
emitted into or absorbed from the heat bath. Again, the real photon
vertex can be either on the $p$ or $p'$ leg, and the contributions of
the two can be independently calculated. The insertion can be a 3-point
vertex (photon inserted on the $p$ or $p'$ leg at a new vertex $\mu$) or
a 4-point vertex (photon inserted on an already existing vertex,
giving seagull but not tadpole diagrams since a real photon is actually
emitted/absorbed).

Unlike the virtual photon insertions, physical momentum is carried away
or brought in by the real photon. Without loss of generality, this can
be accounted for by retaining the momenta of the external scalar legs
to be $p$ and $p'$ and adjusting the momentum at the special vertex $V$
to maintain energy-momentum conservation. Hence the factors are somewhat
different from the virtual photon case: when the $(n+1)^{\rm th}$ photon
is emitted from the $p$ leg, the momentum of the scalar/fermion to the
{\em right} of the insertion $\mu$ is $(p + \sum_{i=1}^q l_i
-k)$ where $q$ is the vertex immediately to the {\em left} of $\mu$;
here $l_i$ are the photon momenta emitted/absorbed at the $i^{\rm th}$
vertex. Similarly, for an emission from the $p'$ leg, the momentum of the
scalar/fermion to the {\em left} of $\mu$ is $(p' + \sum_{i=1}^q
l_i+k)$, where $q$ is the vertex immediately to the {\em right} of
$\mu$. If the photon is absorbed rather than emitted, the sign of $k$
is reversed.

Since the real photon insertions contribute to $\vert {\cal{M}} \vert^2$,
that is, to the cross section, we need to consider thermal modifications
to the phase space. The thermal phase space element corresponding to
the $i^{\rm th}$ real photon with momentum $k_i$ is given by,
\begin{align}
\d\phi_i & = \frac{\d^4k_i}{(2\pi)^4} \, 2\pi \delta(k_i^2) \,
	\left[ \theta(k_i^0) + N(\vert \boldsymbol{k_i} \vert) \right]~.
\label{eq:phase}
\end{align}
Here emission of photons corresponds to $k_i^0 > 0$ and absorption to
$k_i^0 < 0$, thus giving the correct statistical factors of $N\!+\!1$ for
photon emission into, and $N$ for photon absorption from, the heat bath
at temperature $T$. Again, the presence of the thermal number operator
worsens the divergence in the case of real photon emission/absorption
as well, giving a leading IR dependence that is linear, since $N \sim
1/k$ in the soft limit. Note that the presence of the same term acts as
an UV cut off when $k \to \infty$.

We proceed as in GY, re-writing the polarisation sum in the cross
section and separating it into a $\widetilde{K}$ part that potentially
contains the entire IR divergent part and an IR finite $\widetilde{G}$
photon part:
\begin{equation}
\sum_{\rm pol} \epsilon^*_\mu (k)\,\epsilon_\nu (k) = -
g_{\mu\nu}~,
\end{equation}
with
\begin{align} \nonumber
g_{\mu\nu} & = \left\{\strut \left[\strut g_{\mu\nu} -
	\tilde{b}_k(p_f,p_i) k_\mu k_\nu \right] + \left[\strut
	\tilde{b}_k(p_f,p_i) k_\mu k_\nu \right] \right\}~, \nonumber \\
 & \equiv \left\{\left[ \widetilde{G}_{\mu\nu}\right] + \left[
 \widetilde{K}_{\mu\nu}\right]\right\}~,
\end{align}
where the tildes have been used to distinguish the real from the virtual
photon contributions. Since $k^2=0$ for both real photon emission and
absorption, we define,
\begin{equation}
\tilde{b}_k(p_f, p_i)  =  b_k(p_f, p_i)\Big|_{k^2 = 0} =
	\frac{p_f \cdot p_i}{ k\cdot p_f\; k\cdot p_i}~,
\end{equation}
where $p_i$ ($p_f$) corresponds to the initial (final) momentum of the
hard scalar in ${\cal {M}}$ ($\cal{M}^*$) where the real photon of
momentum $k$ is inserted.

\subsubsection{Emission/absorption of real $\widetilde{K}$ photons}
\label{ssec:rkgamma}

The proof that the contribution from $\widetilde{K}$ photons is IR
divergent and can be factored is much simpler than the corresponding
case of virtual photons. The key point to note is that real photons,
whether emitted or absorbed, correspond to thermal type 1 photons, so that
the inserted vertex (either $\mu$ or $\nu$) is of type 1 alone. This is
critical in obtaining a cancellation against the virtual photon
contribution and the significance of this virtual contribution being
proportional to $D^{11}$ alone, as shown in Eqs.~\ref{eq:Kgp'p} and
\ref{eq:Kgpp}, is now clear.

The calculation for photon emission proceeds exactly as in the case
of virtual photon vertex insertion on a $p$ or $p'$ leg (see diagrams
shown in Fig.~\ref{fig:B5}). Again, there is a term-by-term cancellation,
leading to a factor proportional to the matrix element of the $n$ photon
diagram, ${\cal{M}}_n$. Similar insertions on the $p$ leg give a result
proportional to $-{\cal{M}}_n$; the difference in sign with the case
of insertion of $b_k (p',p)$ virtual $K$ photons is due to the fact
that the real photon momentum is always out-going for emitted photons;
while the virtual momentum enters/leaves at the $\nu/\mu$ vertex. The
overall sign is reversed in the case of photon absorption; however,
this is irrelevant and unobserved in the cross section. Adding the two
terms and squaring gives the contribution of the real $\widetilde{K}$
photon insertion to be an overall factor multiplying the $n^{\rm th}$
order cross section, proportional to,
\begin{align} \nonumber
\left \vert {\cal{M}}_{n+1}^{\widetilde{K}\gamma,{\rm tot}} \right \vert^2
	& \propto -e^2 \left[\strut \tilde{b}_k(p,p) -2
	\tilde{b}_k(p',p) + \tilde{b}_k(p',p') \right]~,
		\nonumber \\
   & \equiv -e^2 \widetilde{J}^2(k)~.
\label{eq:ktilde}
\end{align}
The result holds even when some vertices of the lower order graph are
4-point ones, or correspond to virtual photon insertions as well; this
follows from the arguments given for the virtual $K$ photon insertions
in Appendix~\ref{app:vkgamma}.

Before discussing the cross section, we will first complete the discussion
on insertions of real $\widetilde{G}$ photons, which, as expected,
will be IR finite.

\subsubsection{Emission/absorption of real $\widetilde{G}$ photons}
\label{ssec:rggamma}

The proof of IR finiteness of the real $\widetilde{G}$ photon cross
section follows from the same argument as for the virtual $G$-photon
insertion and is not repeated here in detail. Specifically, the case
where the insertions are on different legs ($p'$ and $p$) is relevant
for the real photon insertions. All the cases such as including both 3-
and 4-point vertices, including thermal effects in both photon
and scalar propagators, etc., hold here; there are no tadpole diagrams
in this case and also no quadratic ${\cal O}(k^2)$ contribution that
needs to be cancelled.

The key point to note here is the $k$-dependence of the thermal part.
The leading divergence (logarithmic in the zero temperature case and
linear in the finite temperature case) cancels as before, between the
$g_{\mu\nu}$ and the $\tilde{b}_k k_\mu k_\nu$ parts of $\widetilde{G}$,
owing to the definition of $\tilde{b}_k$.  We are thus concerned
only with terms with powers of $k$ in the numerator which potentially
give logarithmic subdivergences.

The main difference between virtual and real photon insertion is that
the phase space factor is {\em not symmetric} under $k \leftrightarrow
-k$ because of the presence of the theta function, as seen from
Eq.~\ref{eq:phase}. However, the {\em finite temperature part of the
phase space} is symmetric under this exchange since it includes both
photon emission and absorption. These are anyway the only contributions
of interest since any powers of $k$ in the numerator are finite with
respect to the $T=0$ part. This symmetry enables us to symmetrise the
integrand with respect to $k \leftrightarrow -k$ and obtain the
analogous result that real $\widetilde{G}$ photon insertions are IR
finite. Notice that application of the symmetry requires the presence of
both soft photon emission and absorption terms.

Again, the result holds when one of the photons with momentum $k_l$
contributes through its $T=0$ part; in this case, its corresponding
momentum cannot be flipped since its phase space is {\em not symmetric}
under this exchange. We apply the same logic as with skeletal graphs in
the virtual photon case: if this photon is not a part of the controlling
set, there is no divergence associated when it vanishes and this gives
us no trouble. If it is a part of the controlling set, then the
sub-divergence occurs only when all (including this) momenta vanish
simultaneously; however, any power of $k_l$ in the numerator renders the
contribution finite since it contributes through its $T=0$ part and so
again its contribution is finite. The analysis holds when arbitrary
number of these photons contribute through their $T=0$ parts; also when
some of these are virtual photons, since their contribution is always
symmetric in the loop momentum.

Hence, $\widetilde{G}$ photon emissions give a finite contribution to
the cross section.

\subsubsection{The total cross section from real photon
emission/absorption}
\label{ssec:cgamma}

Consider an $n^{\rm th}$ order graph with an arbitrary number of
$\widetilde{K}$ and $\widetilde{G}$ photon insertions. Now $n_K$
$\widetilde{K}$ and $n_G$ $\widetilde{G}$ real photon
emission/absorption can occur
in $n!/n_K!n_G!$ ways; $n=n_K+n_G$, and each real photon carries
away/brings in a
physical momentum $k_l$ from/to the process. Dividing by $n!$ due to $n$
identical photons in the final state, to this order, we have,
\begin{align}
\d\sigma^{\rm real}_n & = \begin{multlined}[t]
	\sum_{n_K=0}^{n} \int \frac{1}{n_K!}
	\left[ \prod_{i=1}^{n_K} \d\phi_i \left\{ -e^2
	\widetilde{J}^2(k_i) \right\} \right] \times
	\frac{1}{n_G!} \left[\prod_{j=n_K+1}^{n} \d\phi_j 
	\left\{ - \widetilde{G}_{\mu\nu} \left \vert {\cal M}^{\mu\nu,
	\tilde{G}\gamma,tot}_{n_G} \right \vert^2 \right\}\right]
	\times \\
  \quad \quad (2\pi)^4 \delta^4\left(p+q-p'-\sum_{l=1}^{n}
  (-1)^{l} k_l \right)~, 
  \end{multlined} 
\label{eq:sigma_real}
\end{align}
where the factor $(-1)^l$ corresponds to $\pm 1$ depending on whether
the photon with momentum $k_l$ is emitted/absorbed. Here the phase space
factor is given by Eq.~\ref{eq:phase} and the factor $\widetilde{J}(k_i)$
contains the IR divergent part. The $k_i$ dependence in the
energy-momentum conserving delta function is removed by the usual trick
of redefining the delta-function:
\begin{equation}
(2\pi)^4 \delta^4\left(p+q-p'-\sum_{l=1}^{n}
	(-1)^l k_l \right) =
	\int \d^4x \exp \left[-i(p+q-p')\cdot x \right]
	\prod_l \exp(\pm i\,k_l \cdot x)~,
\label{eq:delta}
\end{equation}
where the sign of $k_l$ in the last term depends on whether the real
photon was emitted or absorbed; furthermore, we separate out the
$\widetilde{K}$ photon contribution in the last term:
\begin{equation}
\prod_{l=1}^n \exp\left[\pm i\,k_l \cdot x \right] =
	\prod_{k=1}^{n_K} \exp\left[ \pm i\,k_k \cdot x 
	\right] \times
	\prod_{g=n_K+1}^{n} \exp\left[\pm i\,k_g \cdot x \right]~.
\end{equation}
The terms that depend on the $k_k$ $\widetilde{K}$ photon momenta are
then combined with the (common) factor for every $\widetilde{K}$
insertion. Then the total contribution from {\em each} $\widetilde{K}$
photon is:
\begin{equation}
\widetilde{B}(x) = -e^2 \int \widetilde{J}^2(k_k) \d\phi_k
	\exp\left[ {\pm i\, k_k \cdot x}\right]~.
\label{eq:Btilde}
\end{equation}
The total contribution from $\widetilde{K}$ real photons in
Eq.~\ref{eq:sigma_real} can now be factored as,
\begin{align}
\d\sigma^{\rm real, \widetilde{K}}_n & \propto 
	\sum_{n_K=0}^{n} \frac{1}{n_K!}
	\left(\widetilde{B}(x) \right)\strut^{n_K}~,
\label{eq:realK}
\end{align}
and hence can be exponentiated as $n \to \infty$. We will use this factor
and compute the total cross section for the process to all orders.

\subsection{The total cross section to all orders}
\label{sec:ssigmatot}

The all-order corrections to the tree-level cross section for
{$\gamma^{(*)}\phi \to \phi$} arising from both virtual and real (soft)
photon insertions yields the total cross section for this process:
\begin{align} \nonumber
\d\sigma^{\rm tot} & = 
 \begin{aligned}[t]
 \int \d^4x \, e^{-i(p+q-p')\cdot x} \d\phi_{p'}
          \exp\left[\strut B+B^* \right] \exp \left[\widetilde{B}
	  \right] \times 
  \sum_{n_G=0}^{\infty} \frac{1}{n_G!} \\
	  \prod_{j=0}^{n_G} \times
   	\int \d\phi_j e^{\pm i k_j \cdot x}
	  \left[ -G_{\mu\nu} {\cal M}_{n_G}^{\dagger\mu}
	  {\cal M}_{n_G}^\nu \right] ~,
  \end{aligned} \\
 & = \int \d^4x \, e^{-i(p+q-p')\cdot x}\, \d\phi_{p'} 
           \exp\left[ B+B^*+\widetilde{B} \right] \,
	   \sigma^{\rm finite} (x)~,
\label{eq:sigmatot}
\end{align}
where $\sigma^{\rm finite}$ contains the finite $G$ and $\widetilde{G}$
photon contributions from both virtual and real photons. 
The IR divergent parts of both the virtual and real photon
contributions exponentiate and {\em combine to give an IR finite sum},
as can be seen by studying their small-$k$ behaviour:
\begin{align} 
(B+B^*) + \widetilde{B} & = e^2 \int \d\phi_k \left[
	J(k)^2\left\{\strut 1+2N(\vert k^0 \vert)\right\} -
	\widetilde{J}(k)^2 \left\{\strut \left(1+N(\vert k^0
	\vert)\right)e^{i k\cdot x} +
	N(\vert k^0 \vert)e^{-i k\cdot x}
	\right\} \right] \nonumber \\
 & \stackrel{k \to 0}{\longrightarrow} ~0 + {\cal O}(k^2)~.
\label{eq:finite}
\end{align} 
Notice that the cancellation occurs between virtual and real contributions
only when photon absorption terms (last term in Eq.~\ref{eq:finite}
above) are also included. This is the all-order proof of IR finiteness
of the thermal scalar field theory, analogous to that obtained for the
thermal field theory of fermions in Ref.~\cite{Indu}.

%% file: Sections/sec_concl.tex
\section{Summary and Discussion}
\label{sec:concl}

Corrections to typical hard scattering processes from virtual and real
(soft) photon emission combine so that the infra-red (IR) divergences
cancel order by order to all orders in perturbation theory. The IR
finiteness of pure scalar QED at finite temperature was explicitly shown
here to all orders in perturbation theory using the technique of Grammer
and Yennie. The explicit IR finiteness of the corresponding zero
temperature result was also established along the way. Although the IR
behaviour of such theories are expected to be independent of their spin
structure, it was instructive to calculate the details of the scalar
case in order to understand the key role of the (IR finite) 4-point
vertex contributions which enabled the soft terms to be factorised and
exponentiated for the all-order case. In particular, it was shown that
the presence of the IR finite tadpole contributions was crucial in
factorising and exponentiating the IR divergent terms for virtual photon
insertions.

Along with the results of Ref.~\cite{Indu} where the IR finiteness of
thermal fermionic QED was proven to all orders, the present results now
allow the calculation to be extended to the interesting case of the
interaction of dark matter with thermal fermion and scalar fields at
finite temperature. Such results are of importance in precision estimates
of higher order contributions to dark matter interactions such as $\chi
+ {\cal{F}} \to \chi + {\cal{F}}$, where the interaction of the dark
matter particle, $\chi$, with fermions ${\cal{F}}$ is mediated by charged
scalars. Emission and absorption of soft photons from the heat bath in
the early Universe can significantly alter these cross sections. We
address the issue of the IR finiteness of such models of bino-like dark
matter in the companion paper \cite{Pritam}.

%% file: Sections/appa.tex
\section{Feynman rules for scalar QED at finite temperature}
\label{app:Ascalar}

\setcounter{equation}{0}

For convenience, the Feynman rules used in the calculation are listed
here. With the (bosonic) fields $\varphi$ being defined at finite
temperature to satisfy the periodic boundary conditions, namely,
\begin{equation*}
\varphi(t_0) = \varphi(t_0 - i \beta)~,
\end{equation*}
where $\beta = 1/T$, we are faced with field-doubling, where only the
type-1 (physical) component can appear as external legs, while the type-2
(ghosts) may only appear as internal lines. The propagators acquire a
$2 \times 2$ matrix form, with the off-diagonal elements allowing for
conversion of one type into another.

The photon propagator in the Feynman gauge is given by,
\begin{align} \nonumber
i {\cal{D}}_{\mu\nu}^{t_at_b}(k) & = {-g_{\mu\nu}}i{\cal D}^{t_at_b}(k)~,
\nonumber \\
i {\cal D}^{t_at_b}(k) & = \left(\begin{array}{cc} \Delta(k) & 0 \\ 0 &
\Delta^*(k) \end{array} \right) + 2\pi \delta(k^2) N(\vert k^0 \vert ) 
\left(\begin{array}{cc} 1 & e^{\vert k^0 \vert /(2T)} \\
e^{\vert k^0 \vert /(2T)} & 1 
\end{array} \right)~,
\end{align}
where $\Delta(k) = i/(k^2 + i \epsilon)$, and $t_a, t_b (=1,2)$ refer to the
field's thermal type.

The thermal scalar propagator is given by,
\begin{align}
i {\cal{S}}^{t_at_b} (p,m) & = \left( \begin{array}{cc}
\Delta(p) & 0 \\ 0 & \Delta^*(p) \end{array} \right) +
2 \pi \delta(p^2-m^2) N(\vert p^0 \vert) 
\left(\begin{array}{cc} 1 & e^{\vert p^0 \vert /(2T)} \\
e^{\vert p^0 \vert /(2T)} & 1 
\end{array} \right)~,
\end{align}
where $\Delta(p) = i/(p^2-m^2+i\epsilon)$, and $t_a, t_b (=1,2)$ refer to the
field's thermal type. The first term corresponds
to the $T=0$ part and the second to the finite temperature piece; note
that the latter contributes on mass-shell only.

\begin{figure}[htp]
\centering
\includegraphics[width=0.7\textwidth]{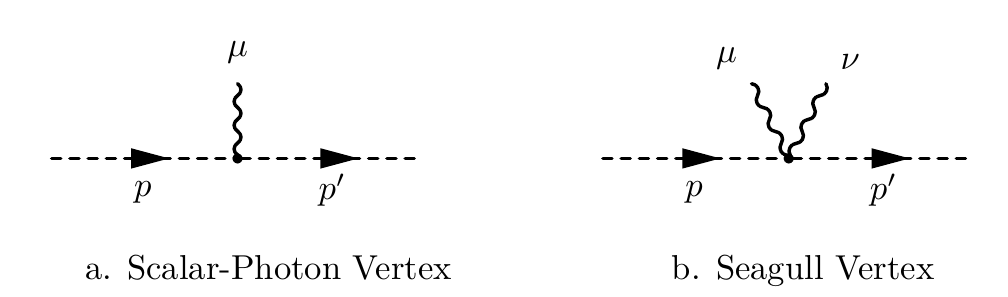}
\caption{Allowed vertices for scalar-photon interactions.}
\label{fig:feynman}
\end{figure}

The scalar--photon vertex factor is $[-ie(p_\mu + p'_\mu)](-1)^{t_a +1}$
where $t_a=1,2$ for the type-1 and type-2 vertices and $p_\mu$ ($p'_\mu$)
is the 4-momentum of the scalar entering (leaving) the vertex. In
addition, there is a 2-scalar--2-photon {\em seagull} vertex (see
Fig.~\ref{fig:feynman}) with factor $[+2ie^2 g_{\mu\nu}](-1)^{t_a+1}$.
All fields at a vertex are of the same type, with an overall sign between
physical (type 1) and ghost (type 2) vertices.

%% file: Sections/appb.tex
\section{Useful identities at finite temperature}
\label{app:fidentities}

\setcounter{equation}{0}

Various identities for scalar fields, useful for simplifying the
calculation, are listed below.

\begin{enumerate}

\item {\bf The propagator}:

The action of $(p^2 - m^2)$ on the scalar propagator is given by,
\begin{align}
(p^2-m^2) ~ i{\cal S}_{\rm scalar~}^{t_at_b}(p,m) & = i (-1)^{t_a+1}
\delta_{t_a,t_b}~.
\end{align}
Henceforth we shall also use the compressed notation, $i{\cal
S}^{t_at_b}(p,m) \equiv i S^{ab}_p$, for convenience.

\item {\bf The generalised Feynman identities}:

Consider an $n^{\rm th}$ order graph with $s$ vertices labelled $s$
to 1 from the hard vertex $V$ to the right (see Fig.~\ref{fig:B1}). We
now insert the $\mu$ vertex of the $(n+1)^{\rm th}$ $K$ photon with
momentum $k$ between vertices $q+1$ and $q$ on the $p'$ leg. Here
the vertex label codes for both the momentum and the thermal type:
the momentum $p'+\textstyle\sum_{i=1}^{q}l_i$ flows to the left of the
vertex $q$ on the $p'$ leg. The photon at this vertex has momentum
$l_q$, with Lorentz index $\mu_q$, and thermal type-index $t_q$. Denoting
$(p'+\textstyle\sum_{i=1}^q l_i)$ as $p'+\Sigma_q$, we have,
\begin{align}
S\strut^{q\mu}_{p'+\Sigma_q}
	\left[(2p'+2\Sigma_q+k)\cdot k\right]
	S\strut^{\mu,q+1}_{p'+\Sigma_q+k}
	& = i (-1)^{t_\mu+1} \left[ S\strut^{q,q+1}_{p'+\Sigma_q}
	\delta_{t_\mu,t_q+1} - S\strut^{q,q+1}_{p'+\Sigma_q+k}
	\delta_{t_\mu,t_q} \right]~.
\end{align}
If the photon vertex is inserted to the right of the vertex labelled
'1' on the $p'$ leg, we have,
\begin{align} \nonumber
[(2p'+k)\cdot k] S\strut^{\mu,1}_{p'+k} & = (-1)^{t_\mu+1}
\delta_{t_\mu,t_1}~,
\end{align}
since $p'^2 = m^2$. Similar relations hold for insertions of vertex $\nu$
of the virtual $K$ photon on the $p$ leg since $p^2 = m^2$ as well.

\end{enumerate}

%% file: Sections/appc.tex
\section{Details of factorisation of virtual $K$ photon insertions}
\label{app:vkgamma}

\setcounter{equation}{0}

\subsection{Both $K$ photon vertex insertions on the $p'$ leg alone:
$T=0$}

Since the simplification and cancellation in this case is non-trivial,
we first focus on the $T=0$ terms alone. The propagator terms simplify to
$1/(P^2 - m^2)$, where $P = p' + {\textstyle\sum}_q$ and is denoted as
$1/(p'+{\mathcal{S}}_q)^2$. We ignore the tadpole contributions for now.

We will consider each set in turn. We start with the Set I terms. As
before, there is a term-by-term cancellation between diagrams with fixed
$\nu$ vertex in Set I, leaving only one term in each such set. The result
from the diagrams in Set I (neglecting an overall factor of $(ie^2)$,
the factor $b(p',p')$, and the loop integration, etc.), again retaining
only the $k_\mu k_\nu$ terms from the photon propagator, is,
\begin{align} 
{\mathcal{M}}_{n+1}^{\mu\nu,p'p', I} & \propto
 \begin{multlined}[t]
	\left\{(2p'+l_1)_{\mu_1} \frac{1}{(p'+\mathcal{S}_1)^2}
	{(2p'+2\Sigma_1+l_2)_{\mu_2}} \cdots
  	\left[\frac{1}{(p'+\mathcal{S}_{s-1})^2}
	(2p'+2\Sigma_{s-1}+l_s)_{\mu_s}-
  	\right. \right. \\
  \left. \left. \frac{1}{(p'+\mathcal{S}_{s-1}+k)^2}
        {(2p'+2\Sigma_{s-1}+2k+l_s)_{\mu_s}} \right]
        \frac{1}{(p'+\mathcal{S}_s)^2} (V) \cdots \right\}+ \\
  \left\{(2p'+l_1)_{\mu_1} \frac{1}{(p'+\mathcal{S}_1)^2}
	{(2p'+2\Sigma_1+l_2)_{\mu_2}} \cdots \left[
	\frac{1}{(p'+\mathcal{S}_{s-2})^2}
	{(2p'+2\Sigma_{s-2}+l_{s-1})_{\mu_{s-1}}} - \right. \right.  \\
  \left. \left. \frac{1}{(p'+\mathcal{S}_{s-2}+k)^2}
	(2p'+2\Sigma_{s-2}+2k+l_{s-1})_{\mu_{s-1}} \right] \cdots
        \frac{1}{(p'+\mathcal{S}_s)^2} (V) \cdots \right\}+ \\
  \left\{\strut  \cdots \right\}+  \\
  \left\{(2p'+l_1)_{\mu_1} \left[\frac{1}{(p'+\mathcal{S}_1)^2}
 	 (2p'+2\Sigma_1+l_2)_{\mu_2}- \right.  \right. \\
  \left. \left. \frac{1}{(p'+\mathcal{S}_1 + k)^2}
  	(2p'+2\Sigma_1+2k+l_2)_{\mu_2}
	  \right] \frac{1}{(p'+\mathcal{S}_2)^2} \cdots
        \frac{1}{(p'+\mathcal{S}_s)^2} (V) \cdots \right\}~.
\end{multlined}
\label{eq:I0}
\end{align}
Similarly, a pair-wise cancellation of terms in Set II occurs, leaving
a single term:
\begin{align}
{\mathcal{M}}_{n+1}^{\mu\nu,p'p', II} & \propto 
 \begin{multlined}[t]
	\left\{(2p'+l_1)_{\mu_1} \frac{1}{(p'+\mathcal{S}_1)^2}
	{(2p'+2\Sigma_1+l_2)_{\mu_2}} \cdots
        \frac{1}{(p'+\mathcal{S}_{s-1})^2}
        (2p'+2\Sigma_{s-1}+l_s)_{\mu_s} \right. \\
 \left. \left[\frac{1}{(p'+\mathcal{S}_s)^2} -
   	\frac{1}{(p'+\mathcal{S}_s+k)^2} \right]
        (V) \cdots \right\}~.
 \end{multlined}
\label{eq:II0}
\end{align}
Set IV has contributions from $(s+1)$ tadpole diagrams; from the
relevant Feynman diagram (see Appendix \ref{app:Ascalar}, it is clear
that virtual $K$ photon tadpole insertions are proportional to a factor
$g^{\mu\nu} k_\mu k_\nu = k^2$ and are hence finite. Let us therefore
first consider the combined contribution of Sets III and IV, excluding
the tadpole contributions. There is both an infrared divergent and a
finite part. Let us first consider the divergent parts. We have 
\begin{align} 
{\mathcal{M}}_{n+1}^{\mu\nu,p'p', III + IV'({\rm div})} & \propto
 \begin{multlined}[t]
  -\left( \left\{(2p'+l_1)_{\mu_1} \frac{1}{(p'+\mathcal{S}_1)^2}
  	{(2p'+2\Sigma_1+l_2)_{\mu_2}} \cdots
        \frac{1}{(p'+\mathcal{S}_{s-1})^2} \right. \right. \times \\
 \left.\left. {(2p'+2\Sigma_{s-1}+l_s)_{\mu_s}}
        \left[ \frac{1}{(p'+\mathcal{S}_s)^2}-
	\frac{1}{(p'+\mathcal{S}_s+k)^2}
	\right] (V) \cdots \right\}\right) \\
% \end{multlined}
% \begin{multlined}[t]
  -\left( \left\{(2p'+l_1)_{\mu_1} \frac{1}{(p'+\mathcal{S}_1)^2}
  	{(2p'+2\Sigma_1+l_2)_{\mu_2}} \cdots
        {(2p'+2\Sigma_{s-2}+l_{s-1})_{\mu_{s-1}}}
	\times \right. \right.  \\
  \left. \left[ \frac{1}{(p'+\mathcal{S}_{s-1})^2}-
         \frac{1}{(p'+\mathcal{S}_{s-1}+k)^2} \right]
        {(2p'+2\Sigma_{s-1}+2k+l_s)_{\mu_s}}
        \frac{1}{(p'+\mathcal{S}_s)^2} (V) \cdots \right\} \\
 +\left\{\strut \cdots \right\}  \\
 +\left\{(2p'+l_1)_{\mu_1} \left[\frac{1}{(p'+\mathcal{S}_1)^2}
	 - \frac{1}{(p'+\mathcal{S}_1+k)^2} \right]
	\times \right. \\
 \left. {(2p'+2\Sigma_1+2k+l_2)_{\mu_2}} \cdots
        \frac{1}{(p'+\mathcal{S}_s)^2} (V) \cdots \right\} \\
 + \left. \left\{ \left(1 - \frac{p'{}^2-m^2}{(p'+k)^2 - m^2} \right) 
         (2p'+2k+l_1)_{\mu_1} 
         \frac{1}{(p'+\mathcal{S}_1)^2} \cdots
	\frac{1}{(p'+\mathcal{S}_s)^2} (V) \cdots \right\}
	\right)~.
 \end{multlined}
\label{eq:34'div}
\end{align}
Here the prime on Set IV$'$ denotes that the tadpole contributions have
not been included. The first term in the round brackets arises from
the first term in Set IV and cancels against the result of Set II while
the second term in round brackets arises from the remaining terms in
Sets III and IV. Here the last term in the second round bracket arises
from self energy corrections to the $p'$ leg and the term proportional
to $(p'^2 - m^2)$ vanishes.

The structure of Eq.~\ref{eq:34'div} (from terms in the second
round bracket alone) is seen to be a sum of terms of the form $\{M_i -
M_j\}$. While this looks very similar to the result from Set I (with an
overall relative negative sign), the second of each term in this set (from
$-M_j$) cancels fully against a similar term in Set I, but the first of
each term (from $M_i$) cancels only partly, leaving behind a tower of
terms with {\em no} $k$ dependence in the denominator, with each term
proportional to $(-2k)_{\mu_q}$, and one additional term, as seen below:
\begin{align}
{\mathcal{M}}_{n+1}^{\mu\nu,p'p', I+II+III+IV'({\rm div})} & \propto
 \begin{multlined}[t]
 \left[ \left\{(2p'+l_1)_{\mu_1} \frac{1}{(p'+\mathcal{S}_1)^2}
 	{(2p'+2\Sigma_1+l_2)_{\mu_2}} \cdots
        \frac{1}{(p'+\mathcal{S}_{s-1})^2} (-2k)_{\mu_s} \right. \right.
	\\
 \left. \frac{1}{(p'+\mathcal{S}_s)^2} (V) \cdots \right\} +
        \left\{\strut  \cdots \right\} \\
 + \left\{(2p'+l_1)_{\mu_1} \frac{1}{(p'+\mathcal{S}_1)^2}
 (-2k)_{\mu_2} \cdots \frac{1}{(p'+\mathcal{S}_s)^2} (V) \cdots \right\} \\
 + \left\{(2p'+l_1)_{\mu_1} (-2k)_{\mu_1}
        \frac{1}{(p'+\mathcal{S}_1)^2} \cdots
        \frac{1}{(p'+\mathcal{S}_s)^2} (V) \cdots \right\} \\
 - \left. \left\{ (2p'+l_1)_{\mu_1} 
         \frac{1}{(p'+\mathcal{S}_1)^2}
	 \cdots \frac{1}{(p'+\mathcal{S}_s)^2} (V)
	 \cdots \right\} \right]~.
\end{multlined}
\label{eq:totdiv}
\end{align}
Here the last two terms arise from the outermost self energy insertions
and the last term is independent of $k$ and is proportional to
$-{\mathcal{M}}_n$. Each of the $s$ terms linear in $k$ are odd under $k
\to -k$ which is allowed under the integral sign and hence vanish, leaving
behind only the term proportional to ${\mathcal{M}}_n$. The finite parts
of Sets III and IV$'$ are given by,
\begin{align}
{\mathcal{M}}_{n+1}^{\mu\nu,p'p', III+IV'({\rm finite})} & \propto
 \begin{multlined}[t]
 \left\{(2p'+l_1)_{\mu_1} \frac{1}{(p'+\mathcal{S}_1)^2}
 	(2p'+2\Sigma_1+l_2)_{\mu_2} \cdots
        (2p'+2\Sigma_{s-1}+l_s)_{\mu_s} \times \right. \\
 \left. \frac{(p'+\mathcal{S}_s+k)^2}{(p'+\mathcal{S}_s)^2}
        \left[\frac{1}{(p'+\mathcal{S}_s)^2}-
	\frac{1}{(p'+\mathcal{S}_s+k)^2} \right]
	(V) \cdots \right\}+ \left\{\strut \cdots \right\} \\
 + \left\{(2p'+l_1)_{\mu_1} \frac{(p'+\mathcal{S}_1+k)^2}
 	{(p'+\mathcal{S}_1)^2} \left[\frac{1}{(p'+\mathcal{S}_1)^2}-
	\frac{1}{(p'+\mathcal{S}_1+k)^2} \right]
	\times \right. \\
 \left. {(2p'+2\Sigma_1+l_2)_{\mu_2}} \cdots
        \frac{1}{(p'+\mathcal{S}_s)^2} (V) \cdots \right\} \\
 + \left[ \frac{(p'+k)^2-m^2}{(p'^2-m^2} -1 \right]
        \left\{(2p'+l_1)_{\mu_1} \frac{1}{(p'+\mathcal{S}_1)^2}
	{(2p'+2\Sigma_1+l_2)_{\mu_2}} \cdots
	\right. \\
 \left. \times \frac{1}{(p'+\mathcal{S}_s)^2} (V) \cdots \right\} ~,
 \end{multlined}
\label{eq:34'finite}
\end{align}
where the last term arises from the self energy correction.  Each of
the finite terms $F_q$ has a $k$ dependence of the form,
\begin{align}
F_q & \sim \cdots
        {(2p'+2\Sigma_{q-1}+l_q)_{\mu_q}}
	\frac{(p'+\mathcal{S}_q+k)^2}{(p'+\mathcal{S}_q)^2} \!
        \left[\frac{1}{(p'+\mathcal{S}_q)^2}-
	\frac{1}{(p'+\mathcal{S}_q+k)^2} \right]
        {(2p'+2\Sigma_q+l_{q+1})_{\mu_{q+1}}} \cdots \nonumber \\
 & = \cdots {(2p'+2\Sigma_{q-1}+l_q)_{\mu_q}}
        \left[\frac{1}{(p'+\mathcal{S}_q)^2} {(2 k \cdot (p'+
	\mathcal{S}_q) + k^2)}
        \frac{1}{(p'+\mathcal{S}_q)^2} \right]
        {(2p'+2\Sigma_q+l_{q+1})_{\mu_{q+1}} }\cdots~,
\label{eq:f_q}
\end{align}	
and hence consists of a term linear in $k$ and one quadratic in $k$.
Note that terms linear in $k$ vanish due to the $k \to -k$ invariance of
the loop integration variable, leaving only terms quadratic in $k^2$
that are IR finite. The requirement for the factorisation and
resummation of the IR divergent terms is that the $K$ photon insertions
be proportional to the lower order matrix element; these additional
finite terms therefore spoil this factorisation process.
The inclusion of the tadpole diagrams precisely cancels these
finite contributions and enables the resummation, as we see below.

We now consider the contribution of the tadpole diagrams in Set IV
(all diagrams with $\nu=\mu$). With the vertex factor now being $i e^2$
rather than $2 i e^2$, the relative weightage between such diagrams and
the corresponding one with two trilinear vertices instead is $1:-1$.
Since each vertex contributes a factor $-g_{\mu\nu} \to -k^2$ for the $K$
photon insertions, it is immediately obvious that the contribution of
the tadpole diagrams is exactly equal and opposite to the finite $k^2$
terms of Sets III+IV$'$, so that these terms cancel as well, leaving
no finite terms. Hence it is important to retain these contributions
while considering the IR behaviour.

% Thus the $(n+1)^{\rm th}$ insertion of a $K$ photon with both vertices
% in the $p'$ leg yields a single term proportional to (the negative of)
% ${\mathcal{M}}_n$ at zero temperature. Now that we understand the
% importance of including the tadpole diagrams, we repeat the calculation
% for the finite temperature case.

\subsection{Both $K$ photon vertex insertions on the $p'$ leg alone:
finite $T$}

We start with the Set I terms. We group sets of diagrams where the $\nu$
vertex is kept fixed, with all possible insertions of the $\mu$ vertex,
with $\mu$ always to the right of $\nu$. There is a term-by-term
cancellation within each of these sub-sets, leaving
only one term in each such set. (Equivalently we can combine diagrams
with fixed $\mu$ vertex.) The result (neglecting overall factors
including $b_k(p',p')$,
and the loop integration, etc., and retaining only the $k_\mu k_\nu$
term in the $K$ photon propagator), is a generalisation of the $T=0$
result in Eq.~\ref{eq:I0}, viz.,
\begin{align}
{\cal{M}}_{n+1}^{\mu\nu,p'p', I} & = \begin{multlined}[t]
 \left\{(2p'+l_1)_{\mu_1} S\strut^{12}_{p'+\sum_1}
	(2p'+2\Sigma_1+l_2)_{\mu_2} \cdots
  	\left[S\strut^{s-1,s}_{p'+\sum_{s-1}}
	(2p'+2\Sigma_{s-1}+l_s)_{\mu_s}
  	\delta_{t_\mu,t_1} \delta_{t_\nu,t_1} 
  	\right. \right. \\
  \shoveright{\left. \left. - S\strut^{s-1,s}_{p'+
  	\sum_{s-1}+k}
        (2p'+2\Sigma_{s-1}+2k+l_s)_{\mu_s}
	\delta_{t_\mu,t_{s-1}} \delta_{t_\nu,t_s} \right]
        S\strut^{s,V}_{p'+\sum_s} (V) \cdots \right\}~~~~~} \\
% {\thinmuskip=0mu\medmuskip=0mu\thickmuskip=0mu
  +\left\{(2p'+l_1)_{\mu_1} S\strut^{12}_{p'+\sum_1}
	(2p'+2\Sigma_1+l_2)_{\mu_2} \cdots \left[
	S\strut^{s-2,s-1}_{p'+\sum_{s-2}}
	(2p'+2\Sigma_{s-2}+l_{s-1})_{\mu_{s-1}}
	\delta_{t_\mu,t_{s-1}} \right. \right.
	% \delta_{t_\nu,t_{s-1}} % }
	\\
   \left. \left. - S\strut^{s-2,s-1}_{p'+\sum_{s-2}+k}
	(2p'+2\Sigma_{s-2}+2k+l_{s-1})_{\mu_{s-1}}
	\delta_{t_\mu,t_{s-2}}\right] \delta_{t_\nu,t_{s-1}} \cdots
        S\strut^{s,V}_{p'+\sum_s} (V)
	\cdots \right\} \nonumber \\
  	\shoveleft{+ \left\{\strut \cdots \right\}}   \\
  	\shoveleft{+ \left\{(2p'+l_1)_{\mu_1} \left[
	S\strut^{12}_{p'+\sum_1}
	(2p'+2\Sigma_1+l_2)_{\mu_2}
	\delta_{t_\mu,t_2} \delta_{t_\nu,t_2} - \right.
	\right.} \\
  \left. \left. S\strut^{12}_{p'+\sum_1 + k}
  	(2p'+2\Sigma_1+2k+l_2)_{\mu_2} \delta_{t_\mu,t_1}
  	\delta_{t_\nu,t_2} \right] S\strut^{23}_{p'+\sum_2} \cdots
        S\strut^{s,V}_{p'+\sum_s} (V) \cdots \right\}. 
 \end{multlined} \\
\label{eq:IT}
\end{align}
The structure is similar to the $T=0$ case, with a generalised form of
the thermal propagators, and the presence of thermal type factors,
$\delta_{t_\mu,t_i} \, \delta_{t_\nu,t_j}$.

Similarly, a pair-wise cancellation of terms in Set II occurs, leaving
a single term:
\begin{multline}
{\cal{M}}_{n+1}^{\mu\nu,p'p', II} = \left\{(2p'+l_1)_{\mu_1}
        S\strut^{12}_{p'+\sum_1}
	(2p'+2\Sigma_1+l_2)_{\mu_2} \cdots
        S\strut^{s-1,s}_{p'+\sum_{s-1}} \delta_{t_\mu,t_s}
        (2p'+2\Sigma_{s-1}+l_s)_{\mu_s} \times \right. \\
 \left. \left[S\strut^{s,V}_{p'+\sum_s} \delta_{t_\nu,t_s} -
 	S\strut^{s,V}_{p'+\sum_s+k} \delta_{t_\nu,t_V} \right]
        (V) \cdots \right\}~.
\label{eq:IIT}
\end{multline}
The contribution from Set III is similar in structure to that from Set I;
we have,
\begin{eqnarray} \nonumber
{\cal{M}}_{n+1}^{\mu\nu,p'p', III} & = & \left\{(2p'+l_1)_{\mu_1}
        S\strut^{12}_{p'+\sum_1}
	(2p'+2\Sigma_1+l_2)_{\mu_2} \cdots
  	\left[S\strut^{s-1,s}_{p'+\sum_{s-1}}
	\delta_{t_\mu,t_s} \right. \right. \nonumber \\
 & &  \left. \left. - 
  	S\strut^{s-1,s}_{p'+\sum_{s-1}+k}
	\delta_{t_\mu,t_{s-1}} \right] \delta_{t_\nu,t_{s-1}}
	(-2k)_{\mu_s} (V) \cdots \right\} \nonumber \\
 & &  +\left\{(2p'+l_1)_{\mu_1}
        S\strut^{12}_{p'+\sum_1}
	(2p'+2\Sigma_1-2k+l_2)_{\mu_2} \cdots \left[
  	S\strut^{s-2,s-1}_{p'+\sum_{s-2}}
	\delta_{t_\mu,t_{s-1}} \right. \right. \nonumber \\
  & & \left. \left. -S\strut^{s-2,s-1}_{p'+\sum_{s-2}+k}
	\delta_{t_\mu,t_{s-2}} \right]
	\delta_{t_\nu,t_{s-1}} (-2k)_{\mu_{s-1}}
        S\strut^{s,V}_{p'+\sum_s} (V) \cdots \right\}
 + \left\{\strut \cdots \rule{0pt}{21pt} \right\} \nonumber \\
 & & + \left\{(2p'+l_1)_{\mu_1} \left[
	S\strut^{12}_{p'+\sum_1} 
	\delta_{t_\mu,t_2} - \right. \right.  \nonumber \\
 & & \left. \left. S\strut^{12}_{p'+\sum_1 + k}
  	\delta_{t_\mu,t_1} \right] \delta_{t_\nu,t_2} (-2k)_{\mu_2} 
	  S\strut^{23}_{p'+\sum_2} \cdots
        S\strut^{s,V}_{p'+\sum_s} (V)
	\cdots \right\} \nonumber \\
 & & + \left\{ \delta_{t_\mu,t_1} (-2k)_{\mu_1} 
 	\delta_{t_\nu,t_1} S\strut^{12}_{p'+\sum_1}
	(2p'+2\Sigma_1-2k+l_2)_{\mu_2}
	\cdots (V) \cdots \right\}~.
\label{eq:IIIT}
\end{eqnarray}
Note that all terms in Eq.~\ref{eq:IIIT} are {\em linearly dependent}
on the inserted photon momentum through the factor $(-2k_{\mu_i})$ and
all but the last term are a set of differences of two terms, of the form
$[T_1-T_2]$. In addition, the $T_1$ terms have {\em no other dependence}
on the momentum of the $K$ photon.

Set IV has two terms per graph, one with $\nu$ inserted immediately to
the left of $\mu$ in all possible ways on the $p'$ leg, and the other a
set of tadpole diagrams with both vertices $\mu$ and $\nu$ being
inserted at the same vertex on the $p'$ leg. A typical term where the
insertion is between $q$ and $q\!-\!1$ vertices on the $p'$ leg gives us,
\begin{multline}
{\cal{M}}_{n+1}^{\mu\nu,p'p', IV,q} = (2p'+l_1)_{\mu_1}
        S\strut^{12}_{p'+\sum_1}
	(2p'+2\Sigma_1+l_2)_{\mu_2}
	\cdots \left\{ \left[S\strut^{q-1,q}_{p'+\sum_{q-1}+k}
	\delta_{t_\mu,t_{q-1}} \delta_{t_\nu,t_q} \right. \right. \\
%   \hspace{1cm} \left. - S\strut^{q-1,q}_{p'+\sum_{q-1}}
 \shoveright{\left. - S\strut^{q-1,q}_{p'+\sum_{q-1}}
	\delta_{t_\mu,t_{q-1}} \delta_{t_\nu,t_{q-1}}
  	+ S\strut^{q-1,\mu}_{p'+\sum_{q-1}}
  	(2p+2\Sigma_{q-1} + k)\cdot k \, \delta_{t_\mu,t_\nu} 
  	S\strut^{\nu,q}_{p'+\sum_{q-1}} \right]} \\
  + \left. \left[S\strut^{q-1,\mu}_{p'+\sum_{q-1}}
  	(-k^2) \delta_{t_\mu,t_\nu} 
  	S\strut^{\nu,q}_{p'+\sum_{q-1}} \right] \right\}~,
\label{eq:IVTq}
\end{multline}
where the terms in the first square bracket come from the first graph
and the term in the second square bracket comes from the second (tadpole)
contribution. It can be seen that the IR finite tadpole contribution is
exactly cancelled by the $k^2$ term from the first graph (last term in
the first square brackets) as was the case at $T=0$. This crucial result
allows the $K$ photon contribution to isolates only the IR divergent
parts. Combining all the graphs, the total contribution to Set IV is,
\begin{eqnarray}
{\cal{M}}_{n+1}^{\mu\nu,p'p', IV} & = & \left\{(2p'+l_1)_{\mu_1}
        S\strut^{12}_{p'+\sum_1}
	(2p'+2\Sigma_1+l_2)_{\mu_2}
	\cdots \left[S\strut^{s,V}_{p'+\sum_s+k}
	\delta_{t_\mu,t_s} \delta_{t_\nu,t_V} \right. \right. \nonumber \\
 & &  \left. \left.  - S\strut^{s,V}_{p'+\sum_s}
	\delta_{t_\mu,t_s} \delta_{t_\nu,t_s}
  	+ S\strut^{s,\mu}_{p'+\sum_s}
  	(2p'+2\Sigma_s)\cdot k \, \delta_{t_\mu,t_\nu} 
  	S\strut^{\nu,V}_{p'+\sum_s} \right]
	\cdots \right\} \nonumber \\
 & &   +\left\{(2p'+l_1)_{\mu_1}
        S\strut^{12}_{p'+\sum_1}
	(2p'+2\Sigma_1+l_2)_{\mu_2}
	\cdots \left[S\strut^{s-1,s}_{p'+\sum_{s-1}+k}
	\delta_{t_\mu,t_{s-1}} \delta_{t_\nu,t_s} \right. \right. \nonumber \\
 & &  \left. \left.  - S\strut^{s-1,s}_{p'+\sum_{s-1}}
	\delta_{t_\mu,t_{s-1}} \delta_{t_\nu,t_{s-1}}
  	+ S\strut^{s-1,\mu}_{p'+\sum_{s-1}}
  	(2p'+2\Sigma_{s-1})\cdot k \, \delta_{t_\mu,t_\nu} 
  	S\strut^{\nu,s}_{p'+\sum_{s-1}} \right] \cdots 
	\right\}
 + \left\{\strut \cdots \rule{0pt}{21pt} \right\} \nonumber \\
 & &   + \left\{(2p'+l_1)_{\mu_1} \left[
        S\strut^{12}_{p'+\sum_1+k} 
	\delta_{t_\mu,t_1} \delta_{t_\nu,t_2} 
        - S\strut^{12}_{p'+\sum_1} 
	\delta_{t_\mu,t_1} \delta_{t_\nu,t_1} \right. \right. \nonumber  \\
 & &  \left. \left. + S\strut^{1,\mu}_{p'+\sum_1}
  	(2p'+2\Sigma_1)\cdot k \, \delta_{t_\mu,t_\nu} 
  	S\strut^{\nu,2}_{p'+\sum_1}
	\right] \cdots \right\} \nonumber \\
 & &  + \left\{(2p')\cdot k \, S\strut^{\mu,1}_{p'} 
 	\delta_{t_\mu,t_\nu} (2p'+l_1)_{\mu_1}
        S\strut^{12}_{p'+\sum_1} 
  	(2p'+2\Sigma_1+l_2)_{\mu_2} \cdots \right. \nonumber \\
 & &  \left. - \delta_{t_\mu,t_1} \delta_{t_\nu,t_1} (2p'+l_1)_{\mu_1}
        S\strut^{12}_{p'+\sum_1}
	(2p'+2\Sigma_1+l_2)_{\mu_2}
	\cdots \right\}~, \nonumber \\
 & &  \equiv \left\{ \left[A_s - B_s + C_s \right] \right\} +
 	\left\{\cdots\right\} +
  	\left\{ \left[ A_1 - B_1 + C_1 \right] \right\} + 
  	\left\{ \left[ -B_0 + C_0 \right] \right\}~.
\label{eq:Mp'p'4}
\end{eqnarray}
Here the last term arises from self-energy corrections on the $p'$
leg, with $B_i$ proportional to ${\cal M}_n$ and all $C_i$ odd in $k$.

The cancellations of various terms are easier to see when added in
pairs. Combining Sets I, II and III, we have,
\begin{eqnarray} \nonumber
{\cal{M}}_{n+1}^{\mu\nu,p'p', I+II+III} & = & \left\{(2p'+l_1)_{\mu_1}
        S\strut^{12}_{p'+\sum_1}
	(2p'+2\Sigma_1+l_2)_{\mu_2}
	\cdots S\strut^{s,V}_{p'+\sum_s} \cdots \times \right.
	\nonumber \\
 & & \left. \left[\strut
	\delta_{t_\mu,t_1}\delta_{t_\nu,t_1} + 
	\delta_{t_\mu,t_2}\delta_{t_\nu,t_2} + \cdots +
	\delta_{t_\mu,t_s} \delta_{t_\nu,t_s} \right]
	\rule{0pt}{21pt}\right\} \nonumber \\
 & & +\left\{\left[(-2k)_{\mu_1} \delta_{t_\mu,t_1} \delta_{t_\nu,t_1}\right]
 	S\strut^{1,2}_{p'+\sum_1}
	(2p'+2\Sigma_1+l_2)_{\mu_2} \cdots +
 	\right. \nonumber \\
 & & + (2p'+l_1)_{\mu_1} S\strut^{1,2}_{p'+\sum_1}
	\left[(-2k)_{\mu_2} 
 	\delta_{t_\mu,t_2} \delta_{t_\nu,t_2} \right]
 	(2p'+2\Sigma_1+l_2)_{\mu_2} \cdots +
	\cdots  \nonumber \\
 & & + \left. (2p'+l_1)_{\mu_1} S\strut^{1,2}_{p'+\sum_1}
 	\cdots S\strut^{s-1,s}_{p'+\sum_{s-1}}
	\left[(-2k)_{\mu_s} \delta_{t_\mu,t_s}
	\delta_{t_\nu,t_s} \right] \cdots \right\} \nonumber \\
 & &  -\left\{(2p'+l_1)_{\mu_1}
        S\strut^{12}_{p'+\sum_1}
	(2p'+2\Sigma_1+l_2)_{\mu_2}
	\cdots \left[S\strut^{s,V}_{p'+\sum_s+k}
	\delta_{t_\mu,t_s} \delta_{t_\nu,t_V} \right]
	\cdots \right. \nonumber \\
 & &  + (2p'+l_1)_{\mu_1}
        S\strut^{12}_{p'+\sum_1}
	(2p'+2\Sigma_1+l_2)_{\mu_2}
	\cdots \left[S\strut^{s-1,s}_{p'+\sum_{s-1}+k}
	\delta_{t_\mu,t_{s-1}} \delta_{t_\nu,t_s} \right]
        \cdots + \cdots \nonumber \\
 & &  + \left. (2p'+l_1)_{\mu_1} \left[
	S\strut^{1,2}_{p'+\sum_1+k}
	\delta_{t_\mu,t_1} \delta_{t_\nu,t_2} \right]
        (2p'+2\Sigma_1+l_2)_{\mu_2} \cdots \right\}~ \nonumber \\
 & & \equiv \left\{\strut X \right\} + \left\{\strut Y \right\} -
 \left\{\strut Z \right\}~.
\label{eq:Mp'p'123}
\end{eqnarray}
Here again, all $Y$ terms are odd in $k$. From a comparison of
Eqs.~\ref{eq:Mp'p'4} and Eqs.~\ref{eq:Mp'p'123}, we see that the $s$
terms labelled $Z$ in Eq.~\ref{eq:Mp'p'123} exactly cancel the $s$
$A_i$ terms in Eq.~\ref{eq:Mp'p'4}. Also, the $s$ terms labelled $X$
in Eq.~\ref{eq:Mp'p'123} exactly cancel the $s$ no of $B_i$ terms ,
$i\ne 0,$ terms in Eq.~\ref{eq:Mp'p'4}. This leaves the set of $C$ and
$Y$ terms in Eqs.~\ref{eq:Mp'p'4} and \ref{eq:Mp'p'123} respectively,
as well as the $B_0$ term in Eq.~\ref{eq:Mp'p'4}.

Now, we have considered the insertion of an $(n+1)^{\rm th}$ virtual
$K$ photon; hence there is an overall integration $\int d^4 k$ which
is symmetric in $(k \leftrightarrow -k)$. So also is the term $b_k$
symmetric, due to its definition, and the photon propagator, ${\cal
D}^{\mu,\nu}_k$ as well. Hence terms odd in $k$ such as the $C$ and $Y$
terms vanish, leaving behind only the $B_0$ term that arose from the
self energy contribution and is proportional to ${\cal M}_n$. Hence the
net $(n+1)^{\rm th}$ virtual $K$ photon contribution is simply $-B_0$,
which is proportional to the lower order matrix element, as was found
for the $T=0$ case.

%% file: Sections/appd.tex
\section{IR finiteness of virtual $G$ photon insertions}
\label{app:vggamma}

\setcounter{equation}{0}

We present here some technical details of $G$ photon finiteness when the
condition of having 3-point vertices in a graph with only a single thermal
photon with momentum $k$, that gives rise to an IR divergence, is
relaxed. We relax the conditions one by one and analyse each in turn.

\begin{enumerate}

\item {\bf Skeletal graphs and virtual $G$  photon insertions}: We
consider only skeletal graphs where the IR divergence occurs only when
each of the controlling momenta, $l_i, i=1,\cdots,m$, simultaneously
vanishes.

Specifically, it was shown in Ref.~\cite{Indu} for the thermal case with
fermionic QED, that symmetrising the $G$ photon integrand with respect
to $(l_i \to -l_i)$ where all the controlling $l_i, i = 1, \ldots,
m$, and $k$, are $G$ photon insertions, results in one or more extra
powers of any of these momenta in the numerator, thus softening the
divergence and removing it. To recap, if only $k$ is the controlling
momentum that determines the thermal logarithmic IR subdivergence, then
the term proportional to ${\cal O}(k)$ is odd in $k$ and vanishes under
symmetrisation $(k \leftrightarrow -k)$; if other photon momenta are
part of the controlling set, the symmetrisation softens and removes this
subdivergence.

The extension of this to the scalar case is straightforward since the
structure of these terms is the same. The thermal part of the photon
propagator is symmetric under $(k \leftrightarrow -k)$ and the above
argument holds. This also trivially extends to the case where some of the
$l_i$ in the controlling set come from $T=0$ contributions which do not
contain $\delta(l_i^2)$ in their propagator terms; this is because the
{\em leading term} still vanishes due to the definition of $b_{l_i}$, and
terms with any power of $l_i$ in the numerator are finite since the $T=0$
part has only a leading log divergence. If the lower order graph contains
$T=0$ photons $l^0_i$ that are {\em not} part of the controlling set,
their propagators are symmetric under $l^0_i \leftrightarrow -l^0_i$
and hence the symmetrisation argument goes through in this case as well.

\item {\bf Including 4-point vertices}: So far we have restricted our
analysis to the case where the $(n+1)^{\rm th}$ photon was only inserted
at new vertices, $\mu$, $\nu$ (3-point vertices), but one or both of
them can be inserted at an already existing vertex to give a 4-point
vertex. This will give rise to terms that have additional factors,
\begin{equation*}
 \left\{S\strut^{q-1,\mu}_{p_f+\sum_{q-1}}
	\left[-2g_{\mu,q}\right] \right\}  
 {\rm and} \left\{S\strut^{m+1,\nu}_{p_i+\sum_{m}+k}
 	\left[ -2g_{\nu,m} \right]\right\}~,
\end{equation*}
within the curly brackets of the two corresponding terms in the LHS of
Eq.~\ref{eq:Gpart}. It was shown in Paper I, that compared to 3-point
vertex insertion, the 4-point vertex has an additional dependence that
is linear in $k$. Hence these contributions can be treated just as
the linearly $k$ dependent terms in the 3-point insertions; hence such
vertices do not affect the result. This can be seen as follows.

Consider an arbitrary $G$-photon insertion as shown in Fig.~\ref{fig:B4},
with vertex $\mu$ inserted either between vertices $q$ and $q\!-\!1$, or at
vertex $q$ on a scalar leg where we ignore (for the present) the thermal
contributions. The relevant part of the combined contribution to the
matrix element reads (where we have not included the contribution from
the photon propagator or the overall loop integration, etc., for the
sake of clarity):
\begin{multline}
M_{n+1,q}^{\mu,G\gamma} \sim \cdots \frac{1}{(p'+\sum_{q-1})^2}
 	\left[ (2p'+2\Sigma_{q-1}+k)_\mu
	\frac{1}{(p'+k+\sum_{q-1})^2}
	(2p'+2\Sigma_{q-1}+2k+l_q)_{\mu_q} \right. \\
	\left. - 2g_{\mu \mu_q} \right]\times 
  \frac{1}{(p'+k+\sum_q)^2}~ \cdots~,
\label{eq:TypeB}
\end{multline}
where $l_q$ is the momentum of the photon inserted at vertex $q$, with
a similar term for insertion of the second $G$-photon vertex, $\nu$,
say at vertex $m$. Factor out the $1/{(p'+k+\sum_{q-1})^2}$
propagator from the first term so that it becomes a multiplying
factor to $-2g_{\mu q}$ in the second term which arises from the Type
B seagull diagrams. Since the $G$-photon is added to a skeletal graph,
the divergence occurs only when all the $l_i$ vanish. Furthermore $p'^2
= m^2$, and so the seagull term reduces to $(2p'\cdot k + k^2)$, so the
leading term is linear in $k$. Hence the seagull terms are linear in $k$
and can be analysed just as these terms were in the discussion above
and shown to be IR finite. The case when the scalar legs are thermal
is discussed below.

Note that this argument does not depend on whether the first vertex at
which the $G$-photon was inserted for the seagull diagram was a $G$- or
$K$-photon vertex. The tadpole diagrams are any way proportional to $k^2$
and hence are IR finite.

\item {\bf Including 4-point vertices in the lower order graph}: So
far we have restricted our analysis to the case where the lower order
graph had only 3-point vertices. Just as with the case of $K$
photon insertion, if one of the vertices of the lower order graph was a
4-point vertex, no new photon (either $K$ or $G$) can be added there;
hence the analysis goes through in the same way for $G$ photon insertion
on a lower order graph with mixed 3- and 4-point vertices.

\item {\bf Scalar lines are thermal}: So far we have considered the
$1/p^2$ part of the scalar propagator in analysing the IR behaviour.
When the scalar field is also thermal, the $G$ photon insertions still
give finite contributions, as discussed in
Section~\ref{ssec:thermal}. We now consider the inclusion of 4-point
vertices.

When we include the 4-point vertices, the trick of combining the $L_0$
propagator with the $g_{\mu,q-1}$ term, as was done in Eq.~\ref{eq:TypeB}
and the text below to show its linear dependence on $k$, cannot be
done since the propagator now has both this as well as a delta function
thermal part. The procedure now is to simplify the product $[g^{\mu\nu}
- b_k k^\mu k^\nu] [{\rm scalar}]_{\mu\nu}$ without substituting for
the scalar propagators. The result is messy and not edifying; in short,
the leading contribution from all terms is of ${\cal O}(1)$ and is
symmetric in $(k \leftrightarrow -k)$; the term with one
4-point vertex insertion has {\em one propagator less} than from inserting
only 3-point vertices and when both the $\mu$ and $\nu$ vertices are of
4-point vertex type, there are {\em two propagators less} than with 3-point
vertex insertions. On counting the overall degree of divergence, we find
that the 3-point vertex insertions have leading linear and subleading
logarithmic divergences as have been discussed above; the terms with
4-point vertex insertions have only logarithmic divergences, and the
one with two 4-point vertex insertions are linear in $k$ and have no
divergence. Hence, the extra divergences arising from one 4-point vertex
insertion have the same behaviour as the subleading terms coming from
4-point vertex insertions alone. These divergences are also removed on
symmetrising the integrand over $( k \leftrightarrow -k)$. When there is
more than one controlling divergence, say, a set $l_i, i = 1,\cdots, m$,
then the analysis can be repeated by determining the divergence when all
these controlling momenta are set simultaneously to zero. In all cases,
the symmetrisation removes the logarithmic subdivergences so that the
the $G$ photon insertions are IR finite.

\item {\bf Including $K$ photons}: If some of the photons
were $K$ photons, we have seen that such insertions reduce to an overall
factor multiplying the lower order graph; this reduction did not depend on
whether the remaining vertices had $K$ or $G$ photon insertions. Hence,
after reduction of all $K$ photons, the matrix element contains only $G$
photon insertions and the above argument goes through; with thermal
scalars as well.

\item {\bf Including real photon vertices}: Finally, if some of the
vertices correspond to real photons, we lose the essential symmetry $(
l_r \leftrightarrow -l_r)$ since real photon emission/absorption contains
a phase space factor $\theta(l_r^0)$/$\theta(-l_r^0)$. The rule is then
to symmetrise the integrand only with respect to virtual momenta; this
yields an IR finite result.

\end{enumerate}

We now have to verify that when we ``flesh out'' skeletal graphs and
include self-energy or other terms, the graph remains IR finite.
As shown in GY, insertions of self energy or vertex corrections are
linear in $k$ and hence IR finite. The argument follows that of GY since
it involves rationalising the denominators and applying the equation of
motion. In addition, we can insert scalar or photon loops on the existing
photon lines in the skeletal graph. Scalar loops do not contribute an
IR divergence due to the presence of the mass term in the propagator;
photon loops are tadpoles whose vertex factors render their contribution
IR finite. Hence the conclusion is not changed when such fleshing out
of skeletal graphs is done for the thermal scalar field theory.